\documentclass{aa}  

\usepackage{graphicx}
\usepackage{txfonts}
\usepackage{comment}
\usepackage{xcolor}
\begin{document}

   \title{The Close AGN Reference Survey (CARS)}

  \subtitle{Tracing the circumnuclear star formation in the super-Eddington NLS1 Mrk~1044}


   \author{N.~Winkel\inst{\ref{MPIA}}
          \and
          B.~Husemann\inst{\ref{MPIA}}
          \and
          T.~A.~Davis\inst{\ref{Cardiff}}
          \and
          I.~Smirnova-Pinchukova\inst{\ref{MPIA}} 
          \and
          V.~N.~Bennert\inst{\ref{CalPoly}}
          \and
          F.~Combes\inst{\ref{Paris}}
          \and
          M.~Gaspari\inst{\ref{INAF},\ref{Princeton}}  
          \and
          K.~Jahnke\inst{\ref{MPIA}}
          \and
          J.~Neumann\inst{\ref{Portsmouth}}
          \and 
          C.~P.~O'Dea\inst{\ref{Winnipeg}}
          \and
          M.~P\'erez-Torres\inst{\ref{Granada},\ref{Zaragoza}}
          \and
          M.~Singha\inst{\ref{Winnipeg}}
          \and 
          G.~R.~Tremblay\inst{\ref{Harvard}}
          \and
          H.~W.~Rix\inst{\ref{MPIA}}
          }

    \institute{
      Max-Planck-Institut f\"ur Astronomie, K\"onigstuhl 17, D-69117 Heidelberg, Germany, \email{winkel@mpia.de} \label{MPIA}
      \and
      Cardiff Hub for Astrophysics Research \&\ Technology, School of Physics \& Astronomy, Cardiff University, CF24 3AA, UK \label{Cardiff}
      \and
      Physics Department, California Polytechnic State University, San Luis Obispo, CA 93407, USA \label{CalPoly}
      \and
      LERMA, Observatoire de Paris, PSL Univ., Coll\`ege de France, CNRS, Sorbonne Univ., Paris, France
      \label{Paris}
      \and
      INAF - Osservatorio di Astrofisica e Scienza dello Spazio, via P. Gobetti 93/3, I-40129 Bologna, Italy
      \label{INAF}
      \and 
      Department of Astrophysical Sciences, Princeton University, 4 Ivy Lane, Princeton, NJ 08544-1001, USA
      \label{Princeton}
      \and
      Institute of Cosmology and Gravitation, University of Portsmouth, Burnaby Road, Portsmouth, PO1 3FX, UK
      \label{Portsmouth}
      \and
      Department of Physics \& Astronomy, University of Manitoba, Winnipeg, MB R3T 2N2, Canada \label{Winnipeg}
      \and
      Instituto de Astrofísica de Andaluc\'{i}a, Glorieta de las Astronom\'{i}a s/n, 18008 Granada, Spain
      \label{Granada}
      \and 
      Departamento de F\'{\i}sica Te\'orica, Facultad de Ciencias, Universidad de Zaragoza, E-50009 Zaragoza, Spain
      \label{Zaragoza}
      \and 
      Center for Astrophysics $|$ Harvard \& Smithsonian, 60 Gardent St., Cambridge, MA 02138, USA
      \label{Harvard}
      }

 
  \abstract
   {The host galaxy conditions for rapid supermassive black hole growth are poorly understood. Narrow-line Seyfert 1 (NLS1) galaxies often exhibit high accretion rates and are hypothesized to be prototypes of active galactic nuclei (AGN) at an early stage of their evolution.}
   {We present VLT MUSE NFM-AO observations of Mrk~1044, the nearest super-Eddington accreting NLS1. Together with archival MUSE WFM data we aim to understand the host galaxy processes that drive Mrk~1044’s black hole accretion.}
   {We extract the faint stellar continuum emission from the AGN-deblended host and perform spatially resolved emission line diagnostics with an unprecedented resolution. Combining both MUSE WFM and NFM-AO observations, we use a kinematic model of a thin rotating disk to trace the stellar and ionized gas motion from 10$\,{\rm kpc}$ down to $\sim$30$\,{\rm pc}$ around the nucleus.}
   {Mrk~1044’s stellar kinematics follow circular rotation, whereas the ionized gas shows tenuous spiral features in the center.
   We resolve a compact star forming circumnuclear ellipse (CNE) that has a semi-minor axis of 306$\,{\rm pc}$. Within this CNE, the gas is metal rich and its line ratios are entirely consistent with excitation by star formation. With an integrated SFR of $0.19 \pm 0.05 \,{\rm M}_\odot {\rm yr}^{-1}$ the CNE contributes 27\% of the galaxy-wide star formation.
   }
   {We conclude that Mrk~1044’s nuclear activity has not yet affected the circumnuclear star formation. Thus, Mrk~1044 is consistent with the idea that NLS1s are young AGN. A simple mass budget consideration suggests that the circumnuclear star formation and AGN phase are connected and the patterns in the ionized gas velocity field are a signature of the ongoing AGN feeding.
   }


   \maketitle
%

\section{Introduction}

AGN accreting with high Eddington ratios are often classified as Narrow-line Seyfert 1 (NLS1) galaxies. 
NLS1s offer an unobscured view of the broad-line region (BLR), such that their optical spectra exhibit both emission lines originating from the high-density BLR and from the low-density narrow-line region (NLR).
As opposed to Broad-line Seyfert 1s (BLS1s), both permitted and forbidden emission lines are narrow (FWHM(H$\beta) <2000\,{\rm km\:s}^{-1}$, \citealt{Goodrich:1989}).
NLS1s often exhibit weak [\ion{O}{iii}] emission (\mbox{[\ion{O}{iii}]/H$\beta$<3}, \citealt{Osterbrock:1985}) and strong \ion{Fe}{ii} emission from the broad line region (BLR). This locates them at the extreme end of the Eigenvector 1 relation \citep[EV1,][]{Boroson:1992,Sulentic:2000} which is believed to be driven by the Eddington ratio.

The relative narrowness of the broad lines in NLS1 is usually interpreted as result of smaller rotational velocities of ionized gas clouds in the BLR. Thus, the inferred central black hole masses of NLS1 are small \mbox{($10^6$ - $10^8$ M$_\odot$}, e.g. \citealt[][]{Zhou:2006,Rakshit:2017}). Although ‘narrow’ broad lines could also be produced by inclined BLR discs in conventional broad line Seyferts 1s \citep[BLS1s,][]{Baldi:2016}, inclination-independent BH mass estimates \cite[e.g.][]{Du:2014,Du:2015,Wang:2014,Pan:2018, Berton:2021} and the host morphologies of NLS1s \citep[][]{Krongold:2001, Jarvela:2017, Jarvela:2018, Berton:2019} suggest genuinely low BH masses.

NLS1s systematically lack large-scale diffuse radio emission \citep{Komossa:2006, Gliozzi:2010, Doi:2013, Lister:2016}, and AGN-ionized gas on kpc scales in contrast to broad-line Seyfert 1s (BLS1s) which suggests little to no AGN activity in their recent past \citep[e.g.][]{Husemann:2008}. It is often speculated that NLS1s represent the early stages of the AGN life cycle where the fuelling of the central super-massive black hole (SMBH) is not yet affected by feedback processes from the host galaxy \citep{Mathur:2000, Collin:2004}.

Although NLS1s share the same classification based on their optical spectra, they exhibit fairly heterogeneous properties.
Only few are detected in the radio among which the dominating emission mechanism ranges from AGN-dominated, composite to host-dominated \citep{Jarvela:2021}. Among the AGN-dominated ones, some show traces of extended jets detected in the radio, reaching distances of tens of kpc from the nucleus.  In individual NLS1 galaxies relativistic jets powered by the central engine have been found \citep{Yuan:2008,Foschini:2011,Foschini:2015} as well as blazar-like phenomena such as high brightness temperature, a double-humped spectral energy distribution or gamma-ray emission \citep{Romano:2018,Paliya:2019,Komossa:2018}. X-ray observations have shown that NLS1s exhibit fast variability on typical time scales of less than one day \citep{Boller:2000} which is expected for the small black hole masses and high accretion rates \citep{Ponti:2012}.

\par

One of the open questions regarding the place of NLS1s among the AGN population is the origin of the exceptionally bright \ion{Fe}{ii} emission from their BLR. The high metallicity is difficult to explain from a galaxy evolution perspective as it requires substantial star formation (SF) to enrich the material in the accretion disk. 
At redshifts around  z$\sim$1 SF galaxies are more likely to host rapidly-growing, radiatively-efficient AGN despite having similar gas fractions and star formation efficiencies as their non-SF counterparts \citep{Nandra:2007, Goulding:2014}. This suggests that both galaxy-scale SF and AGN phase are triggered by the same secular processes.
At intermediate and low redshifts however, the accretion becomes radiatively inefficient and mostly mechanical \citep{Hickox:2009}.
Furthermore, it is matter of discussion how gas is channeled from the galactic scales down to the sub-pc scales of the circumnuclear accretion disc, how the gas transport affects the SF and its spatial distribution in the host, and what is the main driver of such transport (e.g., turbulent mixing, inelastic collisions, gravitational torques, see  \citealt{Gaspari:2020} for a review).

The host galaxies of NLS1s are predominantly classified as disc galaxies and often possess pseudo-bulges \citep{Crenshaw:2003}. They often exhibit enhanced SF in their nuclear regions \citep{Deo:2006} and only a few of them show signs of major mergers or interaction \citep{Ohta:2007}. Therefore, the gas transport towards the galaxy center requires internal process to prolong star formation and the AGN accretion.
This idea is further supported by the frequent presence of pseudo-bulges in NLS1 host galaxies \citep{OrbandeXivry:2011} which are formed from secular processes \citep{Kormendy:2004}. These asymmetries can develop circumnuclear spiral structures \citep{Maciejewski:2004} and enable radial gas migration \citep{Sakamoto:1999,Sheth:2005}.

\par

Local AGN in which circumnuclear star formation has been resolved have low Eddington ratios \citep[e.g.][]{Esquej:2014, Ramos-Almeida:2014, Ruschel-Dutra:2017, Esparza-Arredondo:2018, Knapen:2019} and may therefore have been caused by different fuelling mechanisms than accretion-mode NLS1s. The situation is further complicated by the difficulty to observe close to their unobscured nucleus. It requires an accurate deblending of the AGN emission that is broadened by the instrumental point spread function (PSF) and the faint host emission.
Moreover, the NLS1 population is heterogeneous \citep{Komossa:2006,Jarvela:2017} and low-redshift NLS1s are rare which makes a statistical evaluation of their host galaxy properties challenging.
Given the diversity among NLS1s and their complex behaviour, their classification by their optical classification alone does probably not reflect the underlying physical mechanisms. We therefore need in-detail studies of individual objects to explain their characteristics on the black hole - host galaxy interaction.

\par

Mrk~1044 is a textbook example of a NLS1 and one of the nearest super-Eddington accreting AGN. It is a radio-quiet, luminous ($L_{\rm{bol}} = 3.4\times 10^{44} \, \rm{erg\:s}^{-1} $) NLS1 at $z = 0.0162$ \citep{Husemann:2022}. Its central engine is powered by a black hole with a reverberation mapped mass of $M_\bullet=2.8\times10^6\,\rm{M}_\odot$ \citep{Du:2015}. Numerous studies have confirmed Mrk~1044’s high accretion rate with Eddington fractions $\lambda_{Edd}= L_{bol}/L_{Edd}$ that range from 1.2 \citep{Husemann:2022} up to 16 \citep{Du:2015}. Although Mrk~1044 is a bright X-ray source \citep{Dewangan:2007}, it shows no signs of AGN-driven hot outflows in the form of extended X-ray emission \citep{Powell:2018}. The luminous central region, however, shows frequency-dependent variability and an X-ray excess that is likely caused by relativistic Compton scattering from a high-density accretion disc \citep{Mallick:2018}. In a recent study, \citet{Krongold:2021} used XMM-Newton observations to identify an ultra-fast outflow with a multi-velocity, multi-phase structure originating at the scales of the accretion disk.

On galaxy scales, Mrk~1044 has a barred spiral morphology \citep{Deo:2006} with an estimated \ion{H}{i} mass of $2.6 \times 10^{9}\, \rm{M}_\odot$ \citep{Koenig:2009} and H$_2$ mass of $4\times 10^8 \,\rm{M}_\odot$ \citep{Bertram:2007}.
\cite{Powell:2018} have identified three concentric star forming rings. The outer ring at $\sim 7\,\rm{kpc}$ is likely to be associated with spiral arms in a rotating disk, whereas the inner SF region has not been resolved in earlier studies.
Although Mrk~1044 is an extensively studied nearby NLS1, a complete picture of the host galaxy processes and how they affect its AGN activity is still missing. 
In this work we present adaptive-optics assisted optical integral field spectroscopic (IFU) observations of Mrk~1044 to study the ionized properties and explore the kinematic signatures of the black hole - host galaxy interaction at unprecedented spatial resolution.
Throughout this paper we assume a flat $\Lambda$CDM cosmology with $H_0 = 70 \,\rm{km\:s}^{-1}\rm{Mpc}^{-1}$, $\Omega_M = 0.3$, and $\Omega_\Lambda=0.7$. In this, 1$\arcsec$ corresponds to $0.333\,{\rm kpc}$ at the redshift of Mrk~1044 and the associated luminosity distance is 70.0 Mpc.

\section{Data}

\begin{figure*}
\centering
 \resizebox{\hsize}{!}{\includegraphics{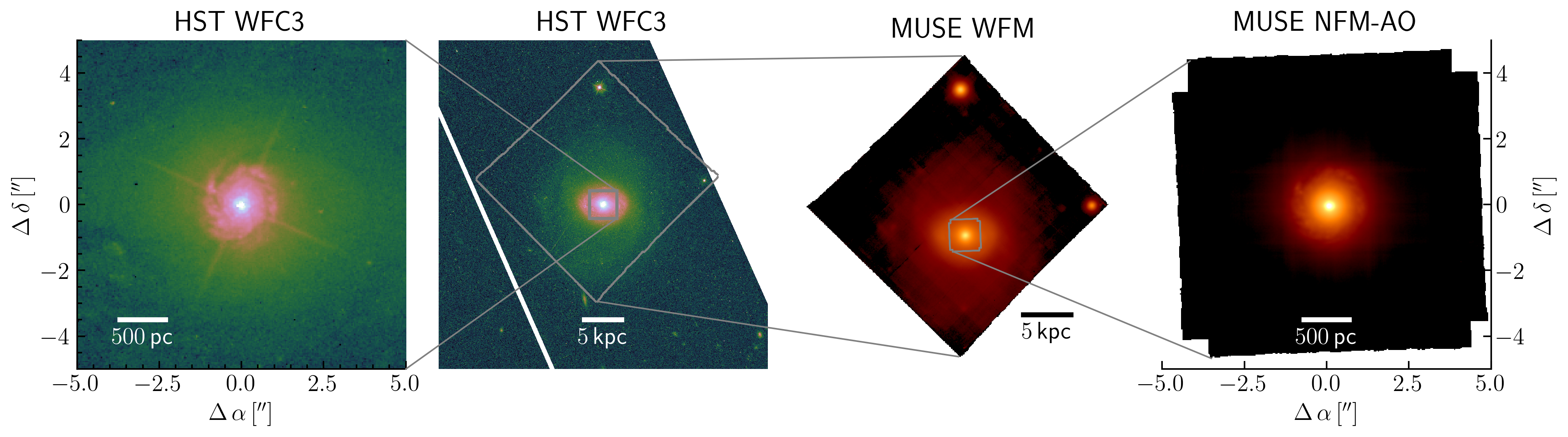}}
 \caption{Comparison of the FOV between Mrk~1044’s observations with HST WFC3/UVIS image in the F547M filter (left), VLT MUSE WFM (center) and VLT MUSE NFM-AO (right). The two images from MUSE are white-light images which were created by integration over the wavelength axis. Both WFC3/UVIS and MUSE WFM have a FOV that covers the host galaxy beyond its effective radius of its disk \mbox{ $r_e = 7.14 \pm 0.04 \,\rm{kpc} $} \citep{Wang:2014}, whereas MUSE NFM-AO only captures the innermost 3.1$\times$3.2$\,\rm{kpc}$.}
 \label{fig:HST_image}
\end{figure*}

\subsection{Observations and data reduction}

For this study we use the integral field spectroscopic data acquired with the Multi Unit Spectroscopic Explorer \citep[MUSE,][]{Bacon:2010,Bacon:2014} at the Very Large Telescope (VLT). The unprecedented spatial resolution and spectral coverage of the adaptive-optics assisted narrow field mode (NFM-AO) allow us to map the host galaxy emission line properties and kinematics of both stellar and gas components down to a resolution of $89\,{\rm mas}$ (30$\,{\rm pc}$). In order to gain a complete picture of the physical processes on different spatial scales, we combine them with the IFU data from the Close AGN Reference Survey \citep[CARS,][]{Husemann:2022} and a high resolution broad band image acquired with the UVIS instrument on the Hubble Space Telescope (HST).

\subsection{Optical Imaging}
Mrk~1044 has been imaged with HST under the program ID 12212 using WFC3/UVIS in the F547M filter. We collect the archival image from the Hubble Legacy Archive\footnote{\url{https://hla.stsci.edu/}}. The field of view (FOV) spans 3.66$\arcmin \times 3.56 \arcmin$ with a spatial resolution of $0\farcs067$ at 6000$\textrm{\r{A}}$ as reported in the WFC3 Instrument Handbook.

\subsection{IFU optical observations}

\subsubsection*{VLT MUSE WFM}
\label{Sect:MUSE-NFM_observations}
As part of the Close AGN Reference Survey \cite[CARS,][]{Husemann:2022}\footnote{\url{https://cars.aip.de/}} Mrk~1044 has been observed with the Wide Field Mode (WFM) of the MUSE instrument. The data were reduced using the standard ESO pipeline v.2.8.1 \citep{Weilbacher:2012,Weilbacher:2020} as described in \cite{Husemann:2022}. The resulting data cube has a FOV of $62\farcs7 \times 63\farcs5 $ with a sampling of $0\farcs2$ per pixel and a seeing-limited resolution of $1\farcs03$. Along the wavelength-axis it extents from $4750\,\textrm{\r{A}}$ to $9300\,\textrm{\r{A}}$ with a spectral resolution of $\sim 2.5\,\textrm{\r{A}}$ that slightly varies over the spectral range \citep{Bacon:2017,Guerou:2017}.

\subsubsection*{VLT MUSE NFM-AO}

\begin{figure}
 \resizebox{\hsize}{!}{\includegraphics{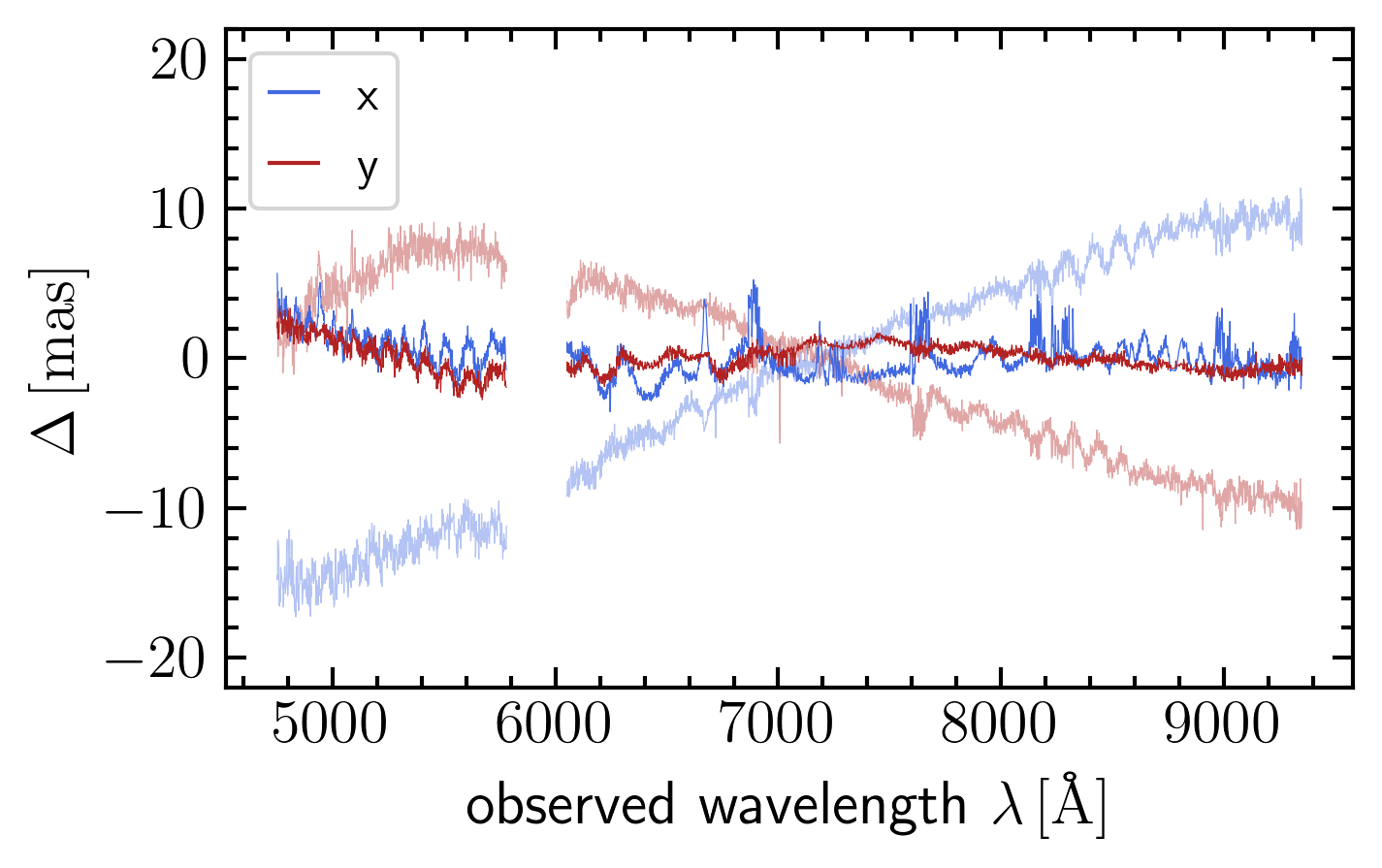}}
 \caption{Wavelength-dependent AGN position on the detector plane before (transparent) and after (opaque) correcting for differential atmospheric refraction. The wavelength range around $5890\,\textrm{\r{A}}$ is blocked by a dichroic in the optical path to avoid contamination and saturation of the detector by the strong sodium emission from the AO laser.}
 \label{fig:DAR_correction}
\end{figure}

\begin{figure*}
   \centering
   \includegraphics[width=\textwidth]{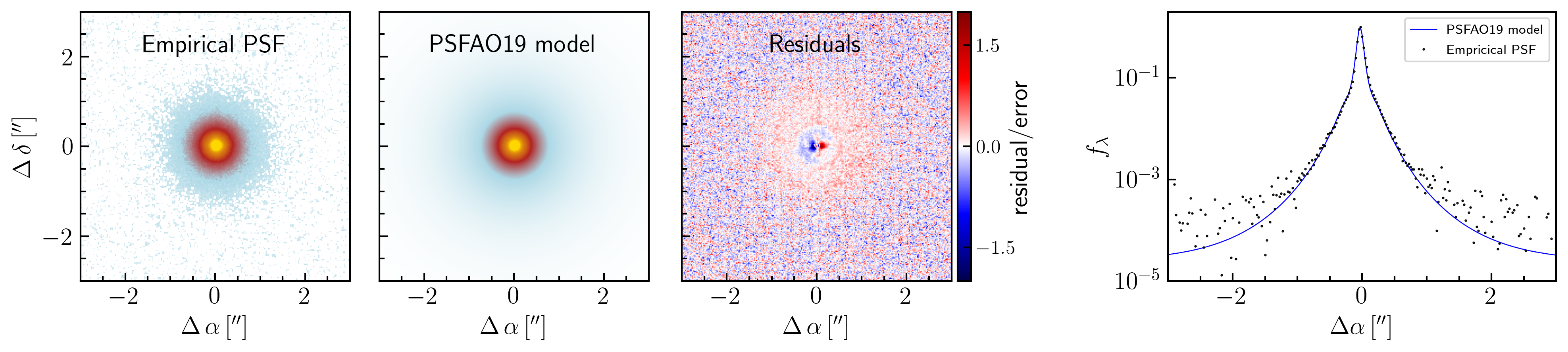}
   \caption{Fitting the MUSE NFM-AO PSF. From left to right the panels show the empirical PSF for the broad H$\beta$ line extracted as described in Sect.~{\ref{sect:Deblending_Broad_Lines}}, the corresponding PSFAO19 model and the residual flux map. The rightmost panel shows the cross section of the PSF at the QSO position and fixed $\delta$. While the turbulent halo of the AO-induced PSF is well reproduced by the PSFAO19 model, the systematic errors near the center are significant compared to the relatively faint host signal.}
              \label{fig:PSFAO19}%
\end{figure*}

We observed Mrk~1044 with the AO-assisted NFM of the MUSE Integral Field Unit under the program 0103.B-0349(A). The data were acquired on 23 Aug 2019, 24 Aug 2019 and 28 Sep 2019. The 12 exposures have an integration time of 550$\,\rm{s}$ each and were dithered by a small angle to minimize the imprint of cosmic rays and flat-fielding artifacts. We use the MUSE pipeline v2.8.3-1 together with the graphical user interface \texttt{ESO Reflex} v2.11.0 to execute the \texttt{EsoRex} Common Pipeline Library reduction recipes.

For WFM data, the reduction pipeline performs a differential atmospheric refraction (DAR) correction that is based on a theoretical prediction \citep{Weilbacher:2020}. However, the correction is poor for observations in the NFM since the performance of the hardware build-in atmospheric dispersion correction (ADC) is insufficient.
In order to correct for the remaining wavelength-dependent atmospheric refraction, we use a 2D Moffat model\footnote{The AO-induced PSF is not well described by a 2D Moffat profile (see Sect.~\ref{sect:PSFAO}), especially at large distance from the center. However, to get an estimate of the data quality it is a sufficiently accurate model.} to measure the AGN position of the individual exposures after the sky correction (\texttt{muse\_scipost}). To remove the wavelength-dependence, we subtract the wavelength-dependent coordinate offset in the PIXTABLE files. Fig.~\ref{fig:DAR_correction} demonstrates the procedure for one example exposure. After the correction, the residual variation of the QSO position across the full wavelength range is of the same order of magnitude as its scatter ($\sim 0\farcs04$).

The spatial resolution of individual exposures depends on the wavelength-dependent performance of the AO system. The science observations taken on 24 Aug 2019 have inferior quality compared to the remaining ones due to the atmospheric conditions during the exposures. Especially at the blue end of the spectrum where we perform most of our diagnostics the difference in resolution is substantial. Here, the FWHM of the 2D Moffat to model the AGN PSF is larger by a factor of $3-4$.
In order to not degrade the resolution of the data cube, we therefore only combine the 8 exposures taken in the night of 23 Aug 2019 and 28 Sep 2019 with \texttt{muse\_exp\_combine}.

The final data cube consists of $369 \times 378$ spaxels, corresponding to a FOV of $9\farcs23 \times 9\farcs45$ with 127,741 spectra. From the width of the telluric emission lines we find a constant resolution of \mbox{${\rm FWHM}=2.54 \pm 0.10\,\textrm{\r{A}}$} across the full wavelength range 4750$\,\textrm{\r{A}}$ - 9350$\,\textrm{\r{A}}$, corresponding to $160.4\,\rm{km\:s}^{-1}$ and $81.5\,\rm{km\:s}^{-1}$ at the blue and the red end of the spectrum respectively.

\subsection{Deblending the AGN and host emission}

In 3D spectroscopic observations of type 1 AGN the light of the unresolved central QSO distributes over the host galaxy as dictated by the PSF. Due to the bright AGN and the low surface-brightness host emission, the AGN contamination is particularly strong within the small FOV of MUSE NFM data cube (see right panel of Fig.~\ref{fig:HST_image}).
Disentangling the host emission from the QSO contamination requires a deblending in both spectral and spatial dimensions. 

The AGN-host deblending process requires five steps:
1) estimation of the  PSF at the broad lines by AGN-host decomposition, 2) reconstructing the AO-shaped PSF at the broad lines with a model, 3) generating a hybrid PSF that combines the empirical PSF with the model PSF, 4) interpolating the hybrid PSF along the wavelength axis, 5) applying an iterative AGN-host deblending procedure combining the wavelength-dependent hybrid PSF with the host galaxy surface brightness profile. In the following subsections, we describe each of the steps in detail.

\subsubsection{PSF extraction at the broad emission lines}
\label{sect:Deblending_Broad_Lines}

In order to separate the host emission from the QSO emission, we employ the dedicated software\footnote{\url{https://git.io/qdeblend3d}} \texttt{QDeblend}$^{\rm 3D}$ \citep[][]{Husemann:2013}. It is based on the concept that the spectrum of each spaxel is a superposition of the host spectrum and a spectrum from the central QSO that is scaled in flux according to the PSF. Since  broad lines from the BLR are spatially unresolved, i.e. point like, type 1 AGN allow to extract the empirical PSF at each broad line available. In the following, we give a brief outline of the concept. For a detailed description of the algorithm we refer the reader to \cite{Husemann:2013}.

As a first step, \texttt{QDeblend}$^{\rm 3D}$ scales the broad-line wings to the flux of the pseudo-continuum to obtain a PSF. The high S/N QSO spectrum is then scaled to match the flux of the broad line wings in each spaxel (step 1), which corresponds to a tentative QSO data cube. As a next step, the emission from the host galaxy is obtained by subtracting the QSO data cube from the original data cube. Since the initially extracted QSO spectrum is inevitably contaminated by a small fraction of host galaxy light this leads to an over-subtraction of the host cube near the center. To estimate the host contribution, \texttt{QDeblend}$^{\rm 3D}$  interpolates the host cube towards the central QSO position assuming a surface brightness (SB) profile. In our case we use a constant SB profile since the over-subtracted region is small ($\sim 0\farcs5$). The host spectrum in the central pixel is then subtracted from the tentative QSO spectrum in order to obtain an updated QSO spectrum that is used for the next iteration starting at step 1.

With this procedure we extracted the empirical PSF at H$\beta$, H$\alpha$ and \ion{O}{i}$\lambda$8446+\ion{Ca}{ii}$\lambda$8498 \citep[see][]{Matsuoka:2007} which have a FWHM of 89, 51 and 41$\,{\rm mas}$ respectively.

\subsubsection{Modelling the MUSE NFM-AO PSF}
\label{sect:PSFAO}

\begin{table*}
\caption{Best-fit PSFAO19 model parameters at three different wavelengths. We extracted the empirical PSFs from the data cube at each of the available broad lines.}             
\label{tbl:PSFAO19_results}      
\centering                          
\begin{tabular}{c c c c c c c c c}        
\hline             
broad line  & $r_0\, [{\rm cm}]$  & $C$ [10$^{-3}$rad$^2$cm$^2$] & A [rad$^2$]& $\alpha$ [cm$^{-1}$] & r & $\theta_R$ & $\beta$  & FWHM$\,[{\rm mas}]$\\    
\hline                        
   H$\beta$ & 13.3 & 14.9 & 337 & 2.36 & 0.987 & 0.35 & 1.56  & 88 \\      
   H$\alpha$ & 18.8  & 10.8 & 705 & 1.34 & 1.008 & -0.44 & 1.63 & 51 \\
   \ion{O}{i}$\lambda$8446+\ion{Ca}{ii}$\lambda$8498  & 26.2 & 7.3 & 8 & 3.28 & 1.018 & -0.61 & 1.67 & 40 \\
\hline                                   
\end{tabular}
\end{table*}

\begin{figure*}
   \centering
   \includegraphics[width=.9\textwidth]{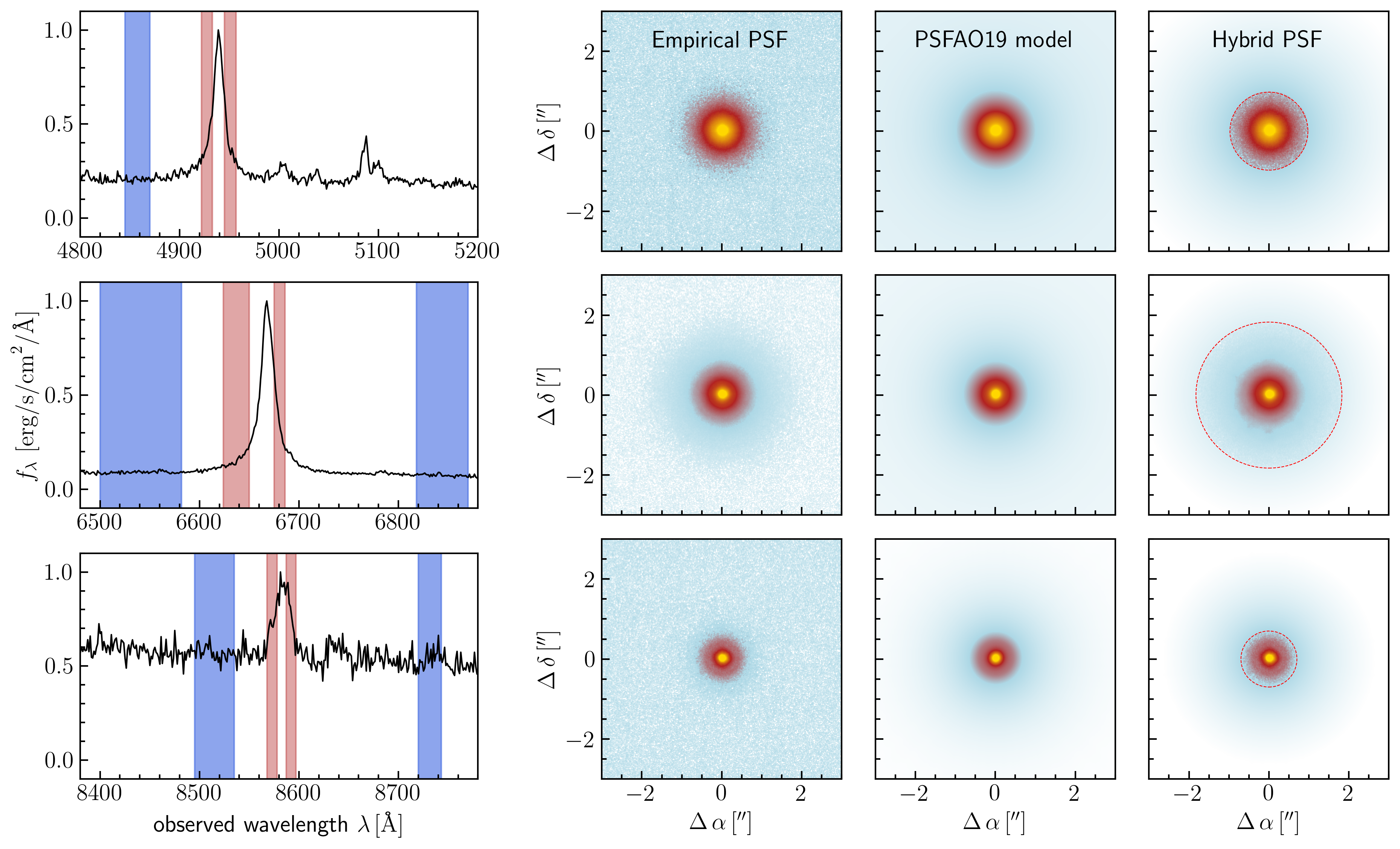}
   \caption{Construction of hybrid PSF from the MUSE data cube. The left panels show the  continuum (blue) and broad line (red) spectral regions from which the broad line intensity is extracted around H$\beta$, H$\alpha$ and \ion{O}{i}$\lambda$8446+\ion{Ca}{ii}$\lambda$8498 (from top to bottom). The resulting empirical PSFs with logarithmic intensity scaling are shown in the 2$^{nd}$ column. The corresponding best-fit PSFAO19 models are shown in the 3$^{rd}$ column. The fourth column shows the hybrid PSFs which we generate by replacing the empirical PSFs with the modeled PSFs beyond a the radius where the S/N drops below a threshold value (dashed red circle).}
              \label{fig:PSFs}%
\end{figure*}

Our observations of Mrk~1044 with MUSE NFM were assisted by the Ground Atmospheric Layer Adaptive Optics for Spectroscopic Imaging (GALACSI) adaptive optics system \citep{Stroebele:2012} to improve spatial resolution. The AO-shaped PSF does not follow the Moffat profile suited for seeing-limited observations. Instead, it has a peculiar shape that depends on the performance of the AO system. Here, we describe the PSF modelling for bright point-like AGN emission required for the subsequent deblending process.

The NFM-AO PSF only slowly changes with wavelength. However, since we subtract the bright AGN emission to get the relatively faint host, already small inaccuracies in the PSF model could severely affect our analysis of the narrow host emission lines. Another problem is that the subtraction of an empirical PSF adds noise to the host data cube which dominates over the faint host emission already at small distances from the center. We are therefore interested in using a model that describes both the behaviour of the NFM-AO PSF towards the faint outskirts and its wavelength dependence. An analytic model for the long-exposure AO-corrected PSF has been presented by \citet[][hereafter PSFAO19]{Fetick:2019}. They describe the phase power spectral density (PSD) with a narrow Moffat core and a wide Kolmogorov halo. The PSFAO19 model consists of 5 parameters (+2 in the asymmetric case) which include the Fried parameter $r_0$, the AO-corrected phase PSD background $C$, the Moffat amplitude $A$ and the Moffat scale factors $\alpha$ and $\beta$. This parametrization is physically motivated by the design of the AO system. The actuators that deform the secondary mirror are separated by a pitch which sets the maximal spatial frequency of the phase $f_{\rm AO}$ that can be corrected by the AO system. Beyond this frequency the residual phase PSD is not affected by the AO system which leaves a residual turbulence in the PSF outskirts. 

To retrieve the best-fit parameters for the empirical PSFs available, we use the PSFAO19 model that uses a Levenberg–Marquardt algorithm to minimize the $\chi^2$-sum of the flux in the residual images. We fit the PSF over the full FOV, weighted by the inverse of the noise variance of the individual pixels. In Table~\ref{tbl:PSFAO19_results} the best-fit parameters are listed. The Fried parameter $r_0$ increases with wavelength as $\lambda ^{1.22 \pm 0.05}$ and is consistent with the theoretical prediction $\lambda ^{6/5}$. Since the atmospheric refraction is smaller in the red (i.e. the AO system performs better), the FWHM of the PSF is decreases with wavelength.

\subsubsection{Generating the hybrid PSF}

In contrast to the empirical PSF, the analytic model has the advantage that it does not contain noise. Thus, it does not add noise to the host cube when subtracting the PSF cube from the original data cube. This is particularly important at large distance from the center where the host emission is faint and the S/N of the empirical PSF is low. We find that the outskirts of the PSF are well reproduced by the PSFAO19 model (see Fig.~\ref{fig:PSFAO19}). Near the center, however, the amplitude of the residuals is $\sim 1$\% of flux from the AGN and therefore of the same order of magnitude as the host signal. 

In order to combine the advantages of both regimes, we create a hybrid PSF. For the high S/N central regions we use the empirical PSF, whereas the model PSF is superior at large distance from the center. We select the transition at the radius beyond which the azimuthally averaged S/N falls below a threshold value of 3. For H$\beta$, H$\alpha$ and \ion{O}{i}+\ion{Ca}{ii} the transition radii are located at 39, 73 and 28$\,{\rm px}$ respectively.

\subsubsection{Interpolation of the PSF with wavelength}

\begin{figure*}
   \centering
   \includegraphics[width=\textwidth]{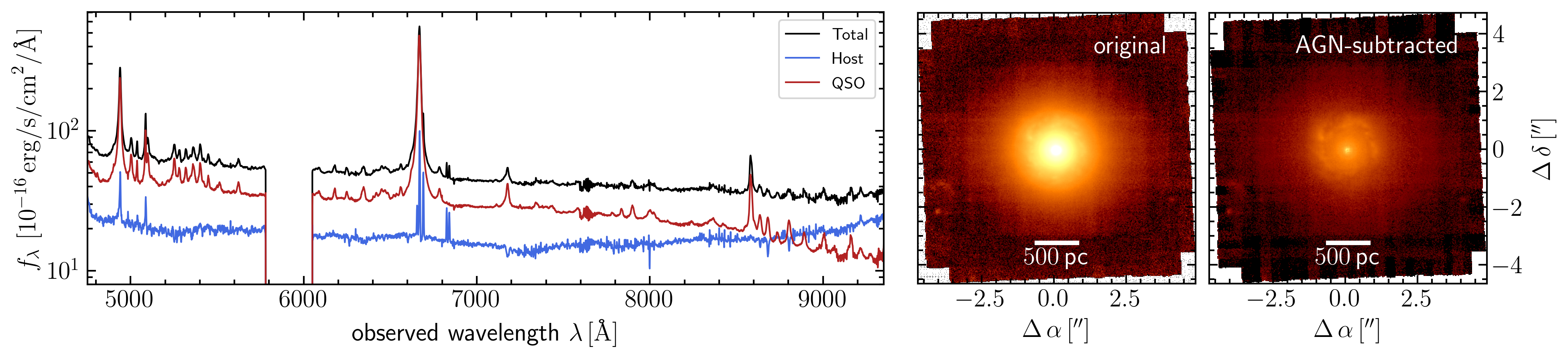}
   \caption{Result of the iterative AGN-host deblending for Mrk~1044 in spectral (left) and spatial (right) dimensions. The panel on the left shows spectrum the original blended spectrum (black) within a 3$\arcsec$ aperture centered on the QSO position. The AGN emission (red) contains the broad lines, whereas in the the host galaxy spectrum (blue) only narrow emission lines are present. 
   The bright point-like AGN emission distributes over the entire FOV of the data cube which is captured by the H$\alpha$ narrow-band images on the right. Both original (left) and deblended host (right) images share the same colour-scaling, which demonstrates the successful deblending.}
              \label{fig:Deblending}%
\end{figure*}

The PSF is wavelength dependent as can be seen from Fig.~\ref{fig:PSFs}. Therefore, we need to interpolate the 2D  PSFs that are available for the three broad lines listed in Table~\ref{tbl:PSFAO19_results}.

As a first approach, we use a spline interpolation of the parameters in Table~\ref{tbl:PSFAO19_results} to generate a PSF cube across the full wavelength range. However, we find that this method leaves strong residual artifacts between the broad lines at which the PSF was extracted. This indicates that not all of the PSFAO19 parameters change smoothly with wavelength and only three broad lines are insufficient to constrain the exact wavelength-dependence. 

An alternative approach involves a simple but effective pixel-by-pixel interpolation of the hybrid PSF across the full wavelength range. Since H$\beta$ and \ion{O}{i}$\lambda$8446+\ion{Ca}{ii}$\lambda$8498 are close to the edges of the MUSE wavelength range, no significant extrapolation is required. As a first step we normalize the hybrid PSFs extracted at H$\beta$, H$\alpha$ and \ion{O}{i}$\lambda$8446+\ion{Ca}{ii}$\lambda$8498 to their peak flux. For each of the spaxels, we describe the wavelength dependence of the flux with a third order polynomial. Consequently, the PSF cube equals the available empirical PSF at the positions of the broad lines. The interpolated PSF cube is then scaled in the central spaxel to match the flux of the QSO spectrum.

\subsubsection{Iterative AGN-host deblending from the IFU data}

We now want to deblend the AGN from the host emission. By construction, the PSF cube contains the faint but non-negligible emission from the host in the central pixel. Therefore, the subtraction from the original datacube causes an over-subtraction in the center. We follow the AGN-host deblending approach described in \cite{Husemann:2022} where the AGN spectrum is iteratively corrected for the host galaxy contribution. Thereby we assume a constant host galaxy surface brightness within the innermost $0\farcs5$ (corresponding to $167\,\rm{pc}$ or $0.08 \,r_e$). 

The result of the iterative deblending process is displayed in Fig.~\ref{fig:Deblending}. The 3$\arcsec$ aperture spectra show that the point-like AGN emission contains the entire broad line emission and little to no contribution from narrow lines. This is because Mrk~1044 AGN is so luminous that it dominates over the narrow line emission on pc-scales. The host spectrum contains the stellar absorption and the spatially extended ionized gas emission in the form of narrow absorption and emission lines respectively. 

Most of our diagnostics in the following are based on emission lines in the H$\beta$-[\ion{O}{iii}] window.
We therefore estimate the spatial resolution for the analysis from PSF cube at H$\beta$ where the ${\rm FWHM}$ is $89\,{\rm mas}$ which corresponds to $30\,{\rm pc}$ in the galaxy system.

\section{Analysis \& Results}

We aim to quantify the stellar and ionized gas properties from the IFU data and map them across the host galaxy. After the AGN-host deblending described in the previous section, the following analysis only employs the AGN-subtracted host emission.

\subsection{Fitting the stellar continuum and the ionized gas emission}
\label{Sect:PyParadise Fitting}

\begin{figure*}
   \centering
   \includegraphics[width=\textwidth]{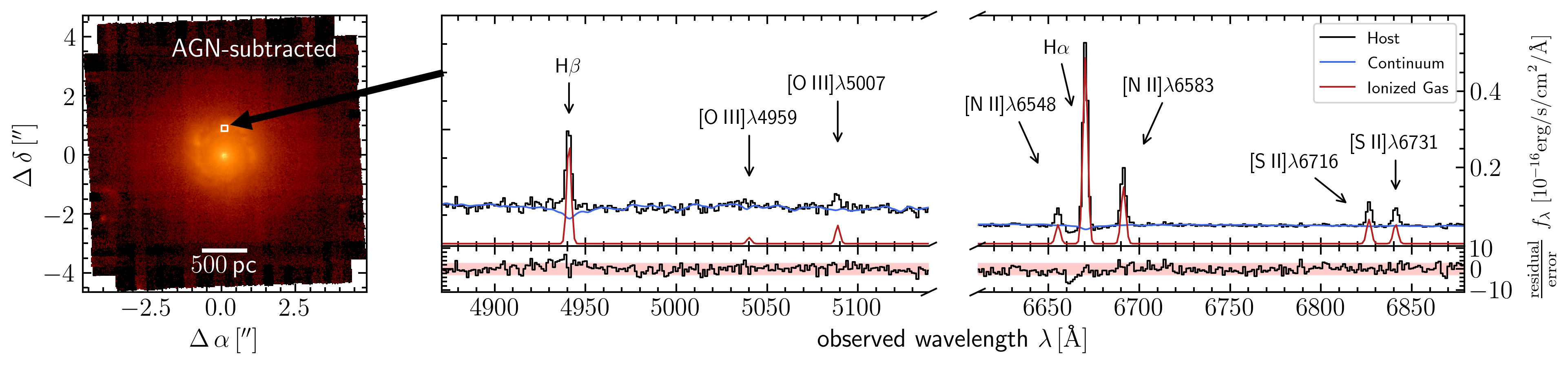}
   \caption{Fitting of an example host spectrum with \texttt{PyParadise}. The panel on the left shows an H$\alpha$ narrow-band image of the AGN-subtracted host emission. An example 8$\times$8-binned spaxel is highlighted with a white square north of the nucleus. The corresponding host spectrum extracted from this aperture is shown in the right panel as a black line in the H$\beta$ (left) and H$\alpha$ window (right). Furthermore, we show the best-fit spectrum of the stellar continuum (blue) and the ionized gas contribution (red) which we obtained with \texttt{PyParadise}. The residuals in the bottom panels show that the model reproduces the spectrum within the 3$\sigma$ confidence region.}
              \label{fig:PyParadise}%
\end{figure*}

For the extraction of the galaxy properties we employ \texttt{PyParadise} \citep{Husemann:2016a, Husemann:2022} which is a publicly available\footnote{\url{https://git.io/pyparadise}} updated version of the stellar population synthesis fitting code \texttt{Paradise} \citep{Walcher2015}. It is extended by an Markov chain Monte Carlo (MCMC) algorithm to find the parameters of the line-of-sight velocity distribution (LOSVD) and a routine to fit the emission lines. The main steps involved are the following.
\par


(1) As a first step, both the input spectra and the template library spectra are normalized by a running mean where emission lines or strong sky line residuals are masked and linearly interpolated. This normalization has the advantage that the fitting result is less sensitive to non-physical continuum variations caused by the wavelength interpolation of the PSF, especially near the nucleus (see Fig.~\ref{fig:Deblending}).

(2) In the next step, \texttt{PyParadise} uses an iterative scheme to independently determine the best-fit of the LOSVD and linear combination of simple stellar population (SSP) template spectra.
Starting from an initial LOSVD guess obtained by fitting a single spectrum drawn from the template library, all spectra in the template library are convolved with the initial LOSVD guess to obtain the best-fit non-negative linear combination of template spectra. 
The resulting best-fit spectrum is used as the input for further iterations and the two step process of estimating the LOSVD and linear combination is repeated. 
For our analysis we select the CB09 SSP library, which is an updated version of the library presented in \cite{Bruzual:2003}.
The template metallicities range from [Fe/H] = -1.44 to 1.44 and stellar ages from $1.7\,{\rm Myr}$ to $13\,{\rm Gyr}$. The SSP spectra cover $3500\,\textrm{\r{A}}$ to $9500\,\textrm{\r{A}}$ at a spectral resolution of $2.5\,\textrm{\r{A}}$. 
Our results do not change within the uncertainties if we employ the higher-resolution SSP library from \cite{Gonzales:2005}.

(3) The final best-fit spectrum is denormalized and subtracted from the original spectrum.

(4) To model the emission lines from the residual spectrum, \texttt{PyParadise} uses a set of Gaussian models with a common line-of-sight velocity, taking into account the spectral resolution that is approximately constant with wavelength (see Sect.~\ref{Sect:MUSE-NFM_observations}). For the doublet emission lines [\ion{N}{ii}]$\lambda\lambda$6548,83 and [\ion{O}{iii}]$\lambda\lambda$4959,5007 we fix the flux ratio to the theoretical prediction of 2.96 \citep{Storey:2000,Dimitrijevic:2007}. Furthermore, we couple the emission lines in radial velocity and velocity dispersion in order to increase the robustness of the flux measurement. Our results do not change within the uncertainties if we do not kinematically tie the model parameters for emission lines with different ionization potentials.

Compared to the established \texttt{pPXF} software \citep{Cappellari:2004,Cappellari:2017} which uses low-order polynomials to fit the stellar continuum, {\rm \texttt{PyParadise} allows us to model the stellar absorption lines irrespective of the details of the PSF subtraction. In this way we can robustly constrain the stellar kinematics close to the nucleus.}
In Fig.~\ref{fig:PyParadise} we visualize an example spectrum from the MUSE data cube within a randomly selected aperture, together with the best-fit spectrum of the stellar continuum and the ionized gas emission.

For a large fraction of the spaxels the S/N is too low to robustly model the low surface brightness emission lines and the faint stellar continuum emission. We therefore employ two spatial binning techniques that spatially co-add the spectra of the host data cube.
For the stellar continuum we use the adaptive Voronoi tessellation routine of \cite{Cappellari:2003} to achieve a minimum signal-to-noise ratio of S/N=20 in the wavelength range 5080$\,\textrm{\r{A}} < \lambda <$ 5260$\,\textrm{\r{A}}$. We then model the binned stellar continuum spectra with \texttt{PyParadise} using the updated CB09 version of the evolutionary synthesis model spectra from \cite{Bruzual:2003} before projecting the stellar kinematics from the Voronoi-grid onto the initial MUSE sampling grid. This information is then used to model the stellar emission again for each spaxel of the host data cube, but this time keeping the stellar LOSVD parameters fixed. Finally, we spatially co-add 2$\times$2 and 8$\times$8 spaxels of the host data cube and fit the emission lines to the residual continuum spectra. The spatial binning allows us to extract the emission lines properties in regions that are more than one order of magnitude fainter than in the original data cube. We estimate the uncertainties for all emission line parameters with a Monte-Carlo approach. Both stellar continuum and emission line fitting are repeated 40 times after each spectrum is fluctuated randomly within the error of each spectral pixel.

\par 

\subsection{Mapping the emission line properties}
\label{sect:eline_maps}

\begin{figure*}
   \centering
   \includegraphics[width=\textwidth]{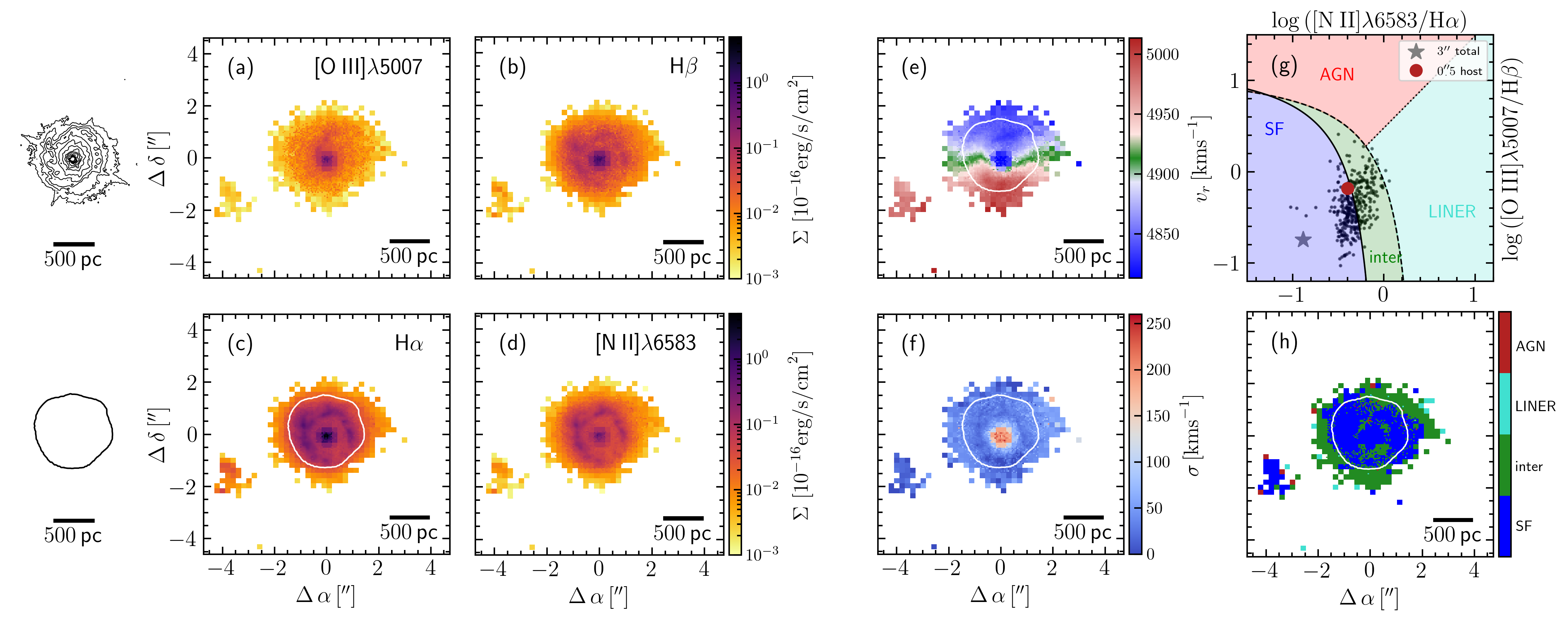}
   \caption{Mapping the ionized gas properties of Mrk~1044.
    The top left panel shows the flux contours of the HST WFC3/UVIS image which capture the spiral dust absorption features. The bottom left panel contains the outer contours where H$\alpha$ surface brightness exceeds $3 \times 10^{-18} \, {\rm erg\:s}^{-1}\textrm{cm}^{-2}$ which indicates the core region of enhanced narrow H$\alpha$ emission. Here and in the following we show both contours as a reference for the size of the structures identified.
    The surface brightness maps for the prominent narrow emission lines after subtracting the AGN emission are shown in (a) - (d). The maps are spatially co-added by 2$\times$2 pixels and 8$\times$8 pixels for regions where S/N of the emission line is $<3$. 
   The LOS velocity (e) and velocity dispersion (f) of Mrk~1044’s ionized gas appear quiescent.
   Panel (g) shows the BPT diagram for Mrk~1044’s central region for 8$\times$8 spatially co-added spaxels. The grey star shows Mrk~1044’s position based on its AGN-contaminated 3$\arcsec$ aperture spectrum. The empirical demarcation lines from  \citealt{Kauffmann:2003} (continuous), \citealt{Kewley:2001} (dashed) and \citealt{Cid-Fernandes:2010} (dotted) define classifications into star forming (SF), composite (inter), Low-Ionization Nuclear Emission Region (LINER) and AGN ionized regions. Already the AGN-blended 3$\arcsec$ spectrum is located in the SF-regime. For the AGN-subtracted cube, all spaxels as well as the 0$\farcs$5-integrated host component are consistent with excitation by star formation. (h) shows the spatial distribution of the excitation mechanism. Mrk~1044 is ionized by star formation, even in the immediate vicinity of its nucleus.}
              \label{fig:eline_maps}%
\end{figure*}

The surface brightness maps together with the velocity field and the dispersion are shown in Fig.~\ref{fig:eline_maps}.
For the surface brightness maps we only select spaxels with S/N>3, and uncertainty of $<30\,\rm{km\:s}^{-1}$ for both velocity and dispersion. For the high S/N regions we overplot the 8$\times$8 binned map with the 2$\times$2 binned spaxels.
The emission from the narrow lines is mostly concentrated in a nearly circular patch around the nucleus. Beyond $\sim$800$\,\rm{pc}$ from the center, the signal from the emission lines is too weak to be robustly modelled. Within the patch, we detect a pronounced ring-like structure that is present in each of the H$\beta$, H$\alpha$ and [\ion{N}{ii}]$\lambda$6583 emission. The ring structure is most prominent in the H$\alpha$ emission for which we extract the contour that is shown in the bottom left panel of Fig.~\ref{fig:eline_maps}.
Since the inclination of the galaxy disk is 67$^\circ$ \citep{Powell:2018}, the deprojected ellipse has an axis ratio of 2.45.
The size of the ring compares to the extent of the luminous inner region in HST WFC3/UVIS image, for which we show the contours in the top left panel of Fig.~\ref{fig:HST_image}.

For each of the surface brightness maps the peak emission stems from near the nucleus. The associated high velocity and the high dispersion in that region indicate an outflowing component.
We note that although the AGN-host deblending algorithm is capable of subtracting any compact emission, it is only a first order subtraction of the point-like emission. As demonstrated by \cite{Singha:2022} the region near the nucleus may require a more complicated modelling with multiple narrow line emitting structures.
Here and in the following analysis, we exclude the emission from the innermost $0\farcs5$ ($160\,{\rm pc}$). A detailed analysis and discussion of Mrk~1044’s outflow will instead be presented in Winkel et al. (in prep.). 
Ignoring the high velocity/dispersion feature in the center, the velocity field exhibits a smooth rotational field across the MUSE NFM FOV. The median dispersion of $\sigma = 34.9\,\rm{km\:s}^{-1}$ is fairly small, indicating dynamically cold ionized gas.

\subsubsection{Excitation mechanism}

From the emission line ratios the underlying excitation mechanism can be identified, i.e. star formation ionization in \ion{H}{ii} regions, photo-ionization by the hard radiation field of an AGN and low-ionization nuclear emission-line regions (LINERs). A commonly used demarcation is made in the Baldwin-Phillips-Terlevic diagram \citep[BPT,][]{Baldwin:1981} that uses the [\ion{O}{iii}]/H$\beta$ vs. [\ion{N}{ii}]/H$\alpha$ line ratios. The upper right panel of Fig.~\ref{fig:eline_maps} shows the BPT diagram for the 8$\times$8 binned spaxels. All spaxels are associated with a cloud that spreads over the SF domain but is elongated towards the AGN ionized region. The scatter originates from the finite spatial resolution of $30\,\rm{pc}$ which does not allow to resolve individual SF clouds. Together with the line-of-sight projection, different ionization conditions are inevitably captured within one spaxel. Furthermore, different ISM metallicities and densities lead to the star forming cloud occupying a large area in the BPT diagram (e.g. \citealt{Smirnova-Pinchukova:2022}). Nonetheless, the vast majority of spaxels in the MUSE NFM FOV is consistent with star formation excitation. We therefore assume the AGN contribution in terms of ionizing radiation to be negligible.

The spatial distribution of the excitation mechanism is shown in the bottom right panel of Fig.~\ref{fig:eline_maps}. Interestingly, the spaxels classified as composite are located towards the outskirts of the bright H$\alpha$ ring. This could either be caused by evolved stars during the short but energetic post-AGB phase which mimic the LINER emission \citep{Binette:1994,Stasinska:2008}, leaking photons from HII regions plus starburst driven shocks  \citep{Heckman:1980, Dopita:1996}.
More strikingly, it becomes evident that even in the very center of Mrk~1044, SF appears to be the dominant excitation mechanism in the host. Since the inner region is dominated by an outflow, we follow the procedure described in \cite{Singha:2022} and use a two-component Gaussian model to fit the central 0$\farcs$5-integrated host spectrum. The resulting BPT classification for the host component is consistent with the single-component analysis of the individual spaxels (see Fig.~\ref{fig:eline_maps}).
This confirms the success of our AGN-host deblending process, since an incorrect treatment of the PSF would generate residual spatially extended AGN ionization structures. On the other hand, it contradicts the findings in nearby AGN where the high excitation is typically found close to the nucleus \citep[][]{Davies:2014,Richardson2014}. 
This is particularly interesting given Mrk~1044’s super-Eddington accretion rate. At such high luminosities quasar mode feedback is expected to drive galactic scale outflows through radiation pressure from the active nucleus \citep{Nesvadba:2007, Liu:2013, Cicone:2018}. However, Mrk~1044’s host does not exhibit any sign of interaction with the AGN photo-ionization field $<1\,{\rm kpc}$ despite the current AGN phase. If its BLR was obscured, Mrk~1044 would  be identified as star forming galaxy. Moreover, the line ratios from the AGN-contaminated 3$\arcsec$-aperture spectrum clearly place Mrk~1044 in the star forming regime (top right panel of Fig.~\ref{fig:eline_maps}). 
There exist several examples of X-ray selected obscured AGN that appear optically inactive \citep[XBONG galaxies, e.g.][]{Severgini:2003,Caccianiga:2007, Trump:2009} and star forming in their diagnostic line ratios \citep{Hornschemeier:2005}. \cite{Castello-Mor:2012} have shown X-ray bright galaxies optically classified as star forming are mostly NLS1s.
This suggests that Mrk~1044 could be a prototype of many accretion-mode AGN at high redshift that are difficult to identify since their spatially unresolved narrow line ratios classify them as star forming galaxies.

\subsubsection{Circumnuclear star formation}
\label{Sect:SFR}

\begin{figure*}
\centering
 \resizebox{.9\hsize}{!}{\includegraphics{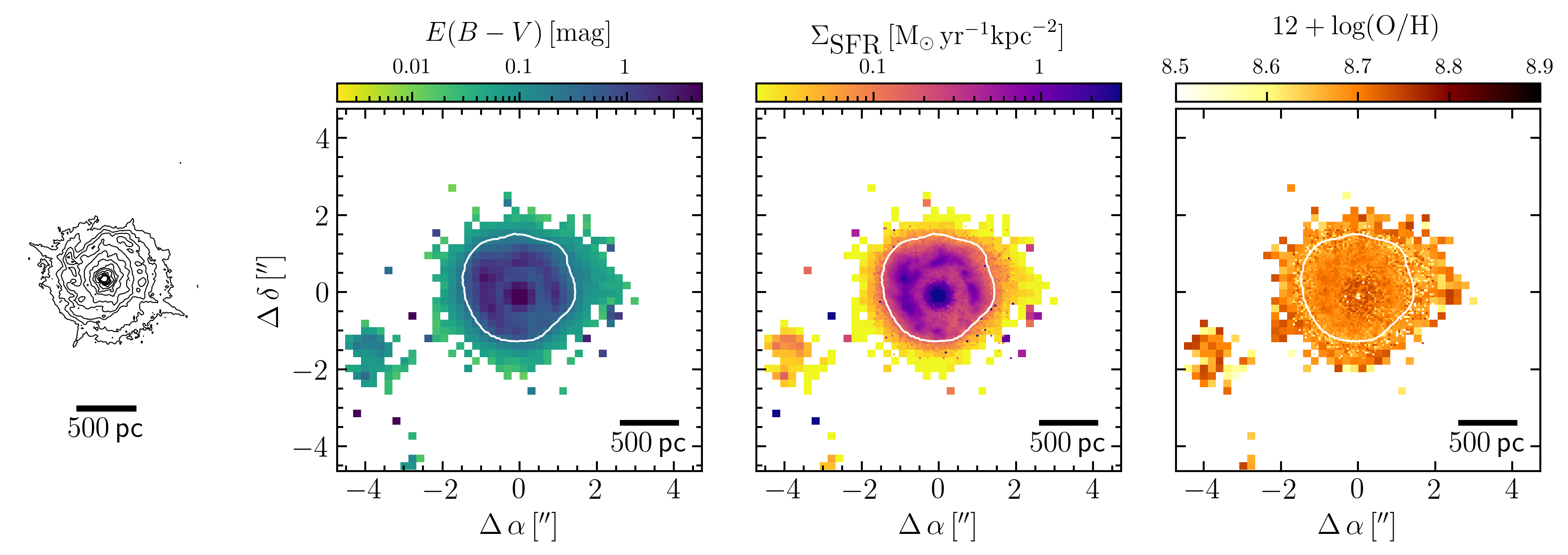}}
 \caption{Mapping the star formation rate in the vicinity Mrk~1044's nucleus. From left to right the panels show (1) the HST/UVIS contours,  (2) Mrk~1044’s extinction map, (3) the star formation rate surface density and (4) the gas-phase metallicity. 
 The contour of the H$\alpha$ ring is over-plotted in white.
 While both extinction and star formation are concentrated in a circumnuclear ring, the chemical composition of the ionized gas is homogeneous within the inner 1$\,{\rm kpc}$.}
 \label{fig:SFR}
\end{figure*}

The star formation rate (SFR) is an important parameter for understanding the interaction between the central AGN and its host galaxy. The surface brightness maps derived in the previous section allow us to map the current star formation rate surface density $\Sigma_{\rm SFR}$ with high spatial resolution.

As a first step we have to correct the H$\alpha$ flux for foreground dust attenuation along the line-of-sight. Assuming case B recombination at an electron temperature $T_e = 10^4 \,\rm{K}$ and electron density $n_e = 100 \, \rm{cm}^{-3}$ the intrinsic Balmer ratio is 2.86 \citep{Osterbrock:1989}. The nebular colour excess can be estimated from the Balmer decrement as
\begin{align}
    E(B-V) 
    &= \frac{E \left( \rm{H}\beta - \rm{H}\alpha \right)}{ \kappa \left( \lambda_{\rm{H}\beta} \right) - \kappa \left(\lambda_{\rm{H}\alpha}\right)} \times \rm{log} \left(\frac{\it F_{\rm{H}\alpha}/ \it F_{\rm{H}\beta}}{2.86} \right) \\
    & = 1.97 \times \rm{log}\left( \frac{\it F_{\rm{H}\alpha}/\it F_{\rm{H}\beta}}{2.86} \right)
\end{align}
Here, $\kappa (\lambda_{\rm{H}\beta})$, $\kappa (\lambda_{\rm{H}\alpha})$ is the reddening curve for star forming galaxies from \cite[][hereafter C00]{Calzetti:2000} evaluated at H$\beta$ and H$\alpha$, respectively.
Using the C00 reddening curve, the dust extinction at wavelength $\lambda$ is connected to the colour excess by
\begin{align}
\label{eq:extinction_law}
A_\lambda & = \kappa \left( \lambda \right) \times E(B-V) \\
A_{\rm{H} \alpha}  &= \left( 3.33 \pm 0.80\right) \times E(B-V)
\end{align}
With the above relation we correct the H$\alpha$ luminosity as
\begin{align}
    L_{\rm intr}({\rm H}\alpha) = L_{\rm obs} ({\rm H}\alpha) \times 10^{\frac{A_{\rm H\alpha}}{2.5}}
\end{align}

Following the argument from the previous section, we assume that all of the extinction-corrected H$\alpha$ luminosity is generated by SF ionization. We then compute the SFR using the relation from \citet[][C07]{Calzetti:2007} which was calibrated for nearby star-forming galaxies assuming a constant SFR over 100$\, \,{\rm Myr}$
\begin{align}
    \left( \frac{\rm SFR}{{[\rm M}_\odot { \rm yr}^{-1}]} \right) = 5.3 \times 10^{-42} \left( \frac{L_{\rm intr} \left( {\rm H}\alpha \right)}{[\rm{erg\:s}^{-1}]} \right)
\end{align}
This calibration applies for the default two power law stellar IMF from Starburst99 \citep{Leitherer:1999}. Moreoever, it contains the underlying assumption that the ISM has solar metallicity.
For a robust measure of the $\Sigma_{\rm SFR}$ we impose a minimum S/N for both H$\beta$ and H$\alpha$ line fluxes. Further, we select $E(B-V)=0$ for spaxels with $F_{{\rm H}\alpha} / F_{{\rm H}\beta} < 2.86$. After applying these criteria the selected $8 \times 8$ binned spaxels sample the entire star forming ring, i.e. we do not lose flux when computing integrated properties of the SF ring.

\par


The left panel of Fig.~\ref{fig:SFR} shows that the extinction is particularly high in the ring of enhanced H$\alpha$ emission.
Within the H$\alpha$-contour (and excluding the innermost $0\farcs5$, see Sect.~\ref{sect:eline_maps}) the median extinction is $0.65\,\rm{mag}$, indicating significant abundance of dust along the line-of-sight. This coincides with the presence of dusty spiral absorption features in the HST WFC3 image (Fig.~\ref{fig:HST_image}).
As shown in the right panel of Fig.~\ref{fig:SFR}, the H$\alpha$-ring shows particularly high SFR surface density. 
The azimuthally averaged peak SFR located at a projected radius of $r=(306 \pm 43)\,\rm{pc}$.
Taking into account the galaxy inclination of 67$^{\circ}$, the deprojected structure is an ellipse with an axis ratio of 2.45. We therefore refer to the circumnuclear region of enhanced SFR as the circumnuclear ellipse (CNE). As before, we exclude the innermost $0\farcs5$ when computing integrated properties of the CNE.

The average $\Sigma_{\rm{SFR}}$ within the CNE is \mbox{$0.16 \pm 0.04 \, \rm{M}_\odot \rm{yr}^{-1}\rm{kpc}^{-1}$}, as computed from the integrated extinction-corrected H$\alpha$ flux. 
This value lies significantly above the minimum of $10^{-3}\, \rm{M}_\odot \rm{yr}^{-1}\rm{kpc}^{-2}$ for driving galactic winds \citep{Veilleux:2005} and is typical of circumnuclear star formation bursts \citep{Kennicutt:1998,Kennicutt:2021}. It is comparable to the lower end of the $\Sigma_{\rm{SFR}}$ in ultra-luminous infrared galaxies (ULIRGS) that host the most powerful starbursts known \citep{Dopita:2002,Genzel:2010}.

For typical star forming spiral galaxies with \mbox{$\Sigma_{\rm gas} = 10 \, {\rm M}_\odot {\rm pc}^{-2}$} the Kennicutt-Schmidt law predicts \mbox{$\Sigma_{\rm SFR} = 6\times10^{-3}\,\rm{M}_\odot \rm{yr}^{-1}\rm{kpc}^{-1}$}.
Following the same procedure as for the CNE, we use the MUSE WFM data cube to compute the galaxy-wide SFR surface density of $\Sigma_{\rm{SFR}} =  (5.9 \pm 1.0) \times 10^{-3} \, \rm{M}_\odot \rm{yr}^{-1}\rm{kpc}^{-1}$. 
We estimate an integrated value of ${\rm SFR_{tot}}= 0.70 \pm 0.17 \, \rm{M}_\odot \rm{yr}^{-1}$. This value is consistent with $ {\rm SFR} = 0.6 \pm 0.2 \,\rm{M}_\odot \rm{yr}^{-1}$ estimated by \cite{Smirnova-Pinchukova:2022}  who used the integrated IR luminosity. While the galaxy-wide value SFR places Mrk~1044 among the star forming population, it is evident that the inner star forming ellipse with $ {\rm SFR_{CNE}}=  0.19 \pm 0.05 \,{\rm M}_\odot {\rm yr}^{-1}$ accounts for 27\% of the absolute galaxy star formation rate.

\subsubsection{Metallicity of the star forming CNE}
\label{sect:metallicity}

Another important characteristic of the ionized gas is the metallicity. In contrast to the star formation rate, which traces short-lived ($\sim$5$\,$Myr) emission from \ion{H}{ii} regions, the gas-phase metallicity provides a cumulative measure for the enrichment history of the ISM. Furthermore, in- and outflows can affect the chemical composition of the gas on short timescales. The gas-phase metallicity therefore provides an important diagnostic to constrain its origin and motion.

There exist several metallicity indicators that are based on the line ratios emitted by \ion{H}{II} regions. Since Mrk~1044’s star forming CNE is completely consistent with excitation by star formation, we can use strong line ratios covered by the MUSE wavelength range to estimate the gas-phase metallicity.
The N2S2 index defined as \mbox{${\rm N2S2} = {\rm log}$ ([\ion{N}{ii}]$\lambda6583$/[\ion{S}{ii}]$\lambda\lambda$6717,31)} has proven to be a reliable indicator for the O/H abundance ratio \citep{Dopita:2016}. 

We use N2S2 index calibration established in \cite{Husemann:2019}. Since the O/H abundance is significantly dependent on the empirical calibration adopted \citep{Kewley:2008}, we use the emission line ratios of star forming galaxies in the 
SDSS DR7 value-added catalog \cite{Brinchmann:2004} to determine a N2S2-calibration
\begin{align}
    12+{\rm log(O/H)} = 8.960 + 0.873 \times {\rm N2S2} - 0.610 \times {\rm N2S2}^2
\end{align}

We thus have a self-consistent measure based on the metallicities from \cite{Tremonti:2004}. Our results do not change within the uncertainties, when we use the \mbox{${\rm O3N2} = {\rm log} \{ ([\ion{O}{iii}] \lambda5007 /{\rm H}\beta) / ([\ion{N}{ii}] \lambda6583 /{\rm H}\alpha) \} $} indicator instead.

The right panel of Fig.~\ref{fig:SFR} shows the gas-phase metallicity of Mrk~1044’s inner region. In contrast to the extinction and the star formation rate surface density, the metallicity does not exhibit a ring-like structure. Within the star forming CNE it is fairly constant with median value of ${\rm 12+log(O/H)} =  9.07$. 
The the scatter of the O/H-abundance ($0.02\,{\rm dex}$) is smaller than the scatter of N2S2-O/H relation ($0.06\,{\rm dex}$), suggesting homogeneous properties of the ISM across the CNE. For the material in the turbulent ISM of the CNE, we estimate an upper limit for the dynamical time by comparing the CNE size $d$ (2$\times$ the semi-minor axis) with the median ionized gas dispersion \mbox{$\sigma_{\rm gas} = 75 \pm 2 \,{\rm km\:s}^{-1}$}
\begin{align}
    t_{\rm dyn}({\rm CNE}) \sim \frac{d}{\sigma_{\rm gas}} = 8\,{\rm Myr}
\end{align}
The dynamical time estimate compares to the timescale of ISM enrichment by massive stars, indicating that the ISM is well mixed on $<1\,{\rm kpc}$ scales. However, the CNE may be affected from gas in- and outflows occurring on longer timescales.
We therefore interpret the constant gas-phase metallicity to be a signature of a homogeneous enrichment process through the ongoing starburst.

\subsection{Kinematic modelling}

\begin{figure}
   \centering
   \includegraphics[width=.7\columnwidth]{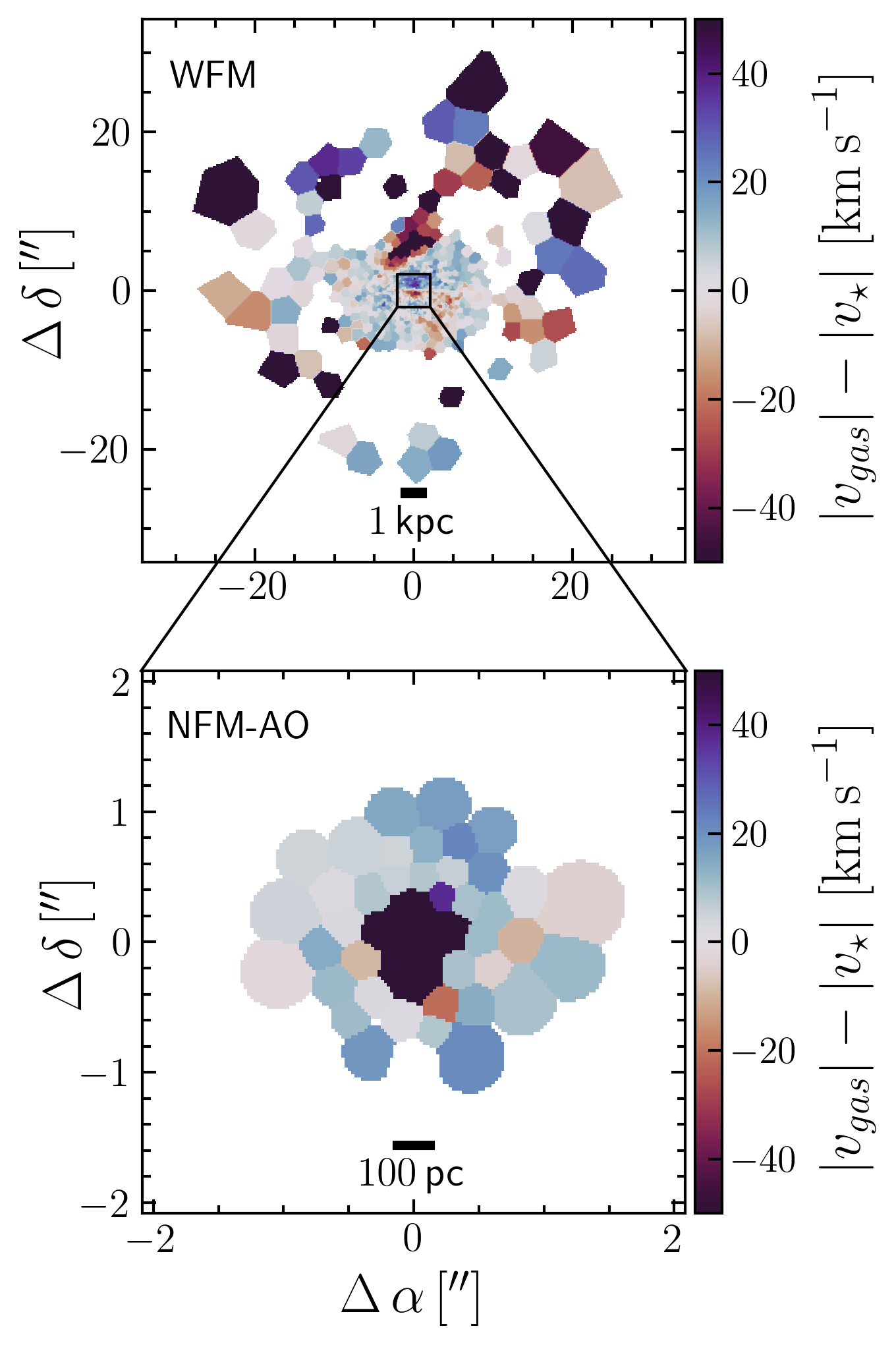} 
   \caption{Difference between Mrk~1044’s gas and stellar velocity field from MUSE WFM (top) and NFM-AO (bottom). Except for the high velocity bloom that stretches $\sim$5$\arcsec$-15$\,\arcsec$ north from the center, the gas rotates faster than the gravitationally driven motion of the stellar component.}
              \label{fig:vstar_vgas}%
\end{figure}

\begin{figure*}
   \centering
   \includegraphics[width=\textwidth]{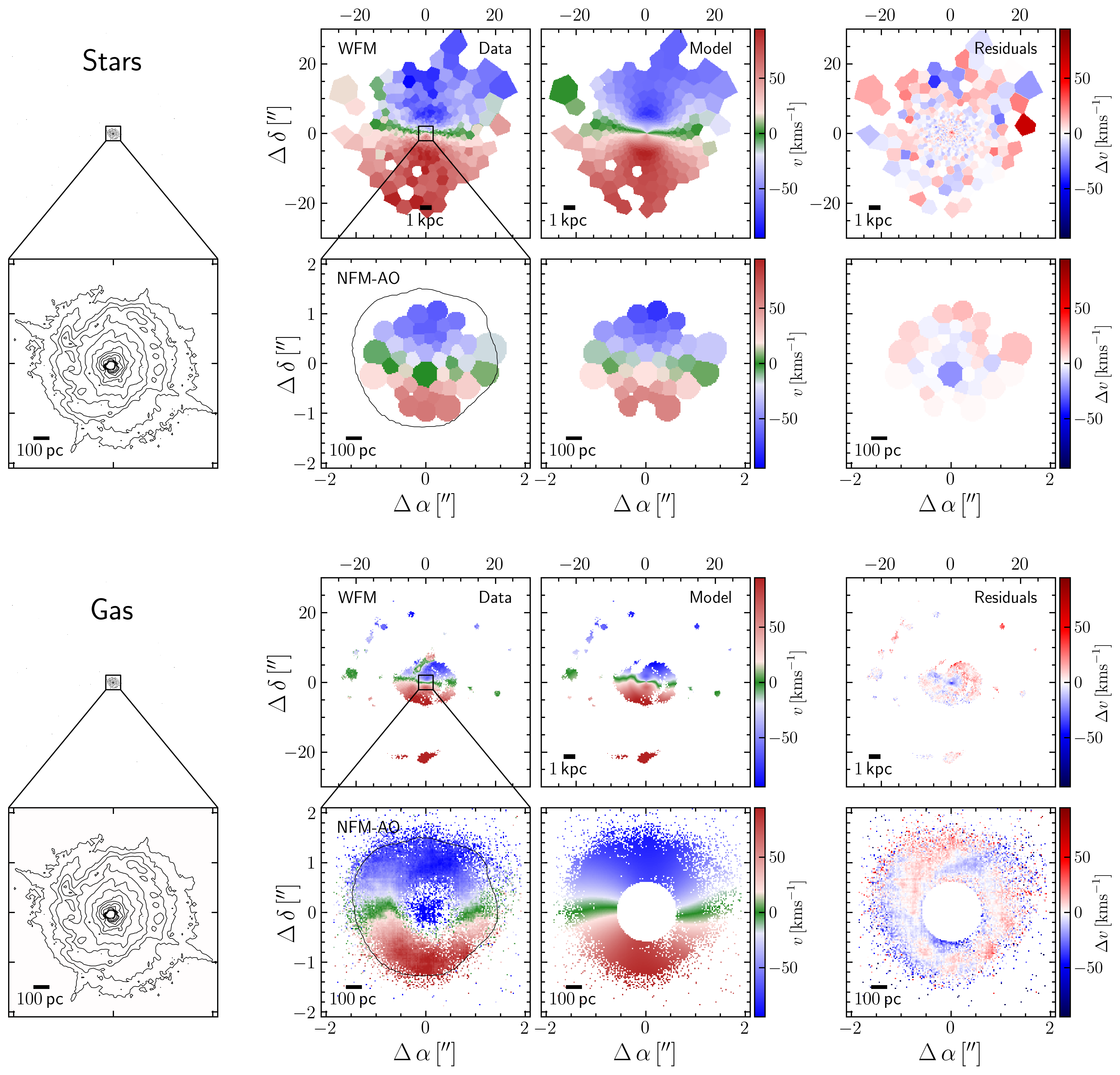} 
   \caption{Modelling the stellar (top) and ionized gas velocity field (bottom) with a thin rotating disk. For each of the two the panels are organized as following:
   From left to right the panels show (1) the contours of the HST WFC3/UVIS image (2) Mrk~1044’s stellar/ionized gas velocity field (3) the kinematic model of a rotating disk and (4) the difference between the observed and model velocity field. For each stars and ionized gas respectively, the top panels show the MUSE WFM data and the bottom panels the zoom-in onto the NFM-AO data. For the ionized gas velocity field we exclude the innermost $0\farcs5$ where the gas does not co-rotate with the disk.
   Both the stellar and ionized gas are well reproduced by the rotating disk model. Only the ionized gas shows tenuous clumpy structures in the velocity residuals which are highlighted in Fig.~\ref{fig:arc}.
   }
              \label{fig:kin}%
\end{figure*}

To investigate whether the gas and stellar motion are consistent with dynamically cold rotation on a circular orbit, we use a simple model to fit the 2D velocity maps extracted in Section~\ref{Sect:PyParadise Fitting}. 
Let’s consider the first moment of the LOSVD, the intensity-weighted mean velocity, which we assume to trace the material in a thin disk. If the inclination of the system $i$ is unknown, the projected velocity of a circular orbit in the plane of the galaxy can be expressed through the azimuthal angle $\phi$ measured from the projected major axis in the galaxy plane:
\begin{align}
v_r 
& = v_{sys} + v \sin i \times \cos \left( \phi + \rm{PA} \right) \\
& = v_{sys} + v \sin i \times \cos \left( \arctan \left( \frac{y-y_c}{x-x_c} \right) + \rm{PA} \right)
\end{align}
Here, $v_{sys}$ is the systemic velocity, PA is the position angle of the rotation axis and $(x,y)$ are the Cartesian coordinates in the observed plane with the galaxy center located at $(x_c,y_c)$.
We assume the AGN position to be the same as the kinematic center of the circular motion. Furthermore, we adopt the inclination of the disk $i = 67.4 \pm 4.5 ^{\circ}$ computed by \cite{Powell:2018}.
Thus, our analytic model consists of three free parameters ($v_{sys},{\rm PA}, v$) for which we determine the best-fit values using the Bayesian MCMC software \texttt{GAStimator}\footnote{\url{https://github.com/TimothyADavis/GAStimator}}. This algorithm maximises the log-likelihood of the residuals between the observed and the model LOSVD.
In order to trace the rotation curve $v(r)$, we sample the FOV with concentric rings within which the model is evaluated. We chose their radial width to equal the FWHM of the PSF, i.e. $1\farcs03$ for MUSE WFM and 89$\,{\rm mas}$ for NFM-AO respectively.
Finally, we assume a smooth rotation curve by radially interpolating the parameters (PA, $v$) to compute the model velocity in each pixel.

\subsubsection{Kinematics of the stellar component}
\label{Sect:kin_model_stars}

Although the stars are likely to have significant pressure support, we conduct a simple experiment here by assuming purely circular motions. The stellar emission is spatially binned such that the stellar continuum reaches S/N$\geq$20 as described in Sect.~\ref{Sect:PyParadise Fitting}. Since the bin size varies across the FOV, we discretize the model by projecting it onto the Voronoi grid and taking the average velocity within each bin. We then use this information to maximize the log-likelihood of the LOSVD. 

The top panels of Fig.~\ref{fig:kin} shows the stellar velocity field together with the best-fit model and the residuals. Both observations from MUSE WFM and NFM-AO show a smooth rotational pattern that ranges from the smallest resolved scales of $\sim$100$\,{\rm pc}$ to galaxy scales $\sim$20$\,{\rm kpc}$. The thin rotating disk model reproduces this pattern well as we do not see systematic extended structures in the residual map. This indicates that the stellar velocity field is largely unperturbed, justifying our initial (simplistic) assumption of circular orbits for the stellar rotation.
Approaching the center, the stars have decreasing rotational velocity $v$, while the PA stays constant. This is compatible with the quiescent rotating velocity field of a disk galaxy that is dominated by the gravitational potential of the host galaxy alone.

\subsubsection{Kinematics of the ionized gas phase}

\begin{figure}
   \centering
   \includegraphics[width=\columnwidth]{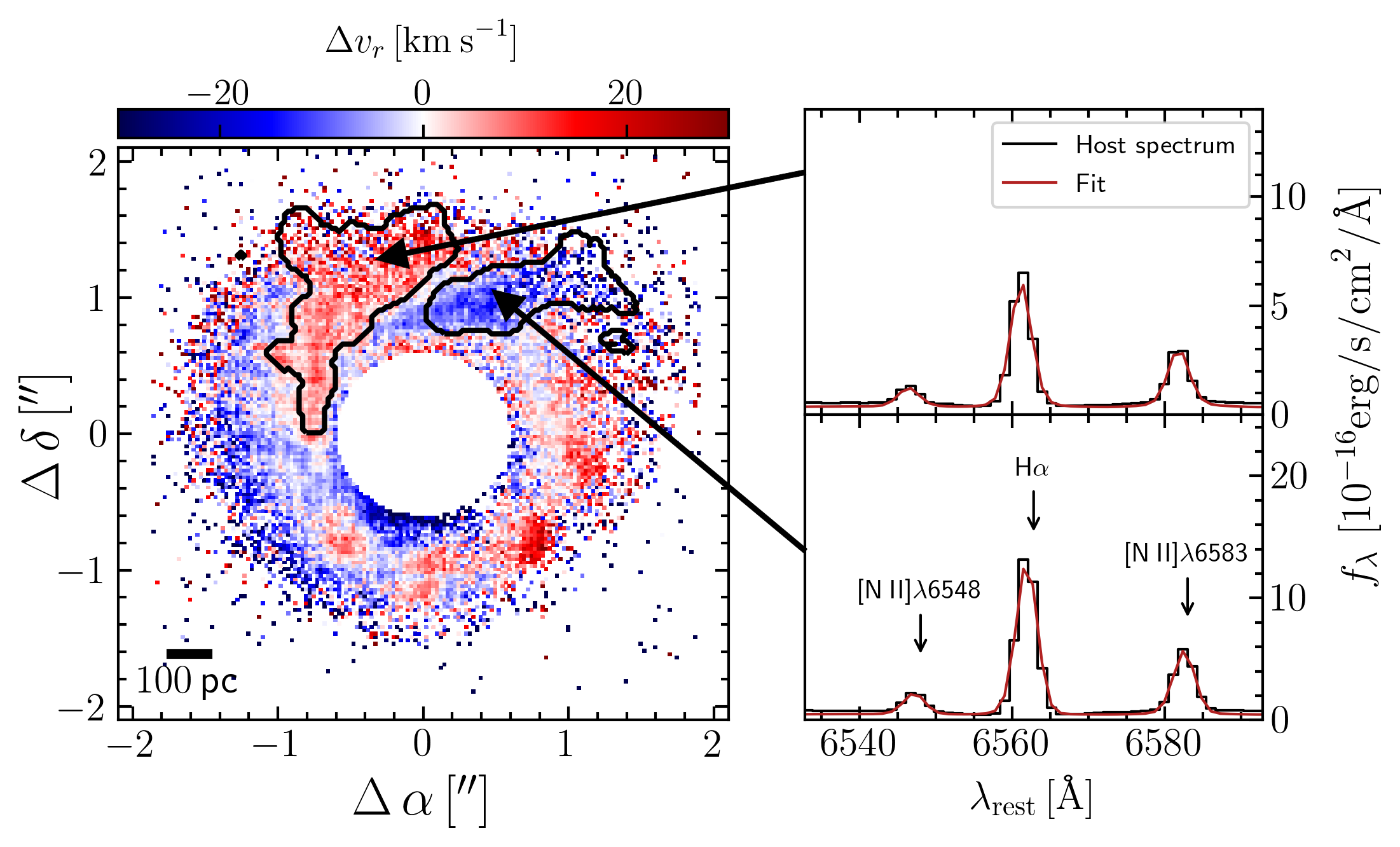} 
   \caption{Zoom-in onto the clumpy structures detected in the residual gas velocity $\Delta v_r$ after subtracting the rotating thin disk model. The contours highlight spaxels of the smoothed map (see text) and correspond to an approaching (blue) and receding (red) clump respectively. The panels on the right show the rest-frame spectra of the host in the H$\alpha$+[\ion{N}{ii}] region, integrated over the respective patch. 
   The emission lines have a symmetric shape and are well reproduced by the fit, indicating a genuine shift of the velocity residuals $\Delta v_r$.}
              \label{fig:arc}%
\end{figure}

\begin{figure}
   \centering
   \includegraphics[width=.8\columnwidth]{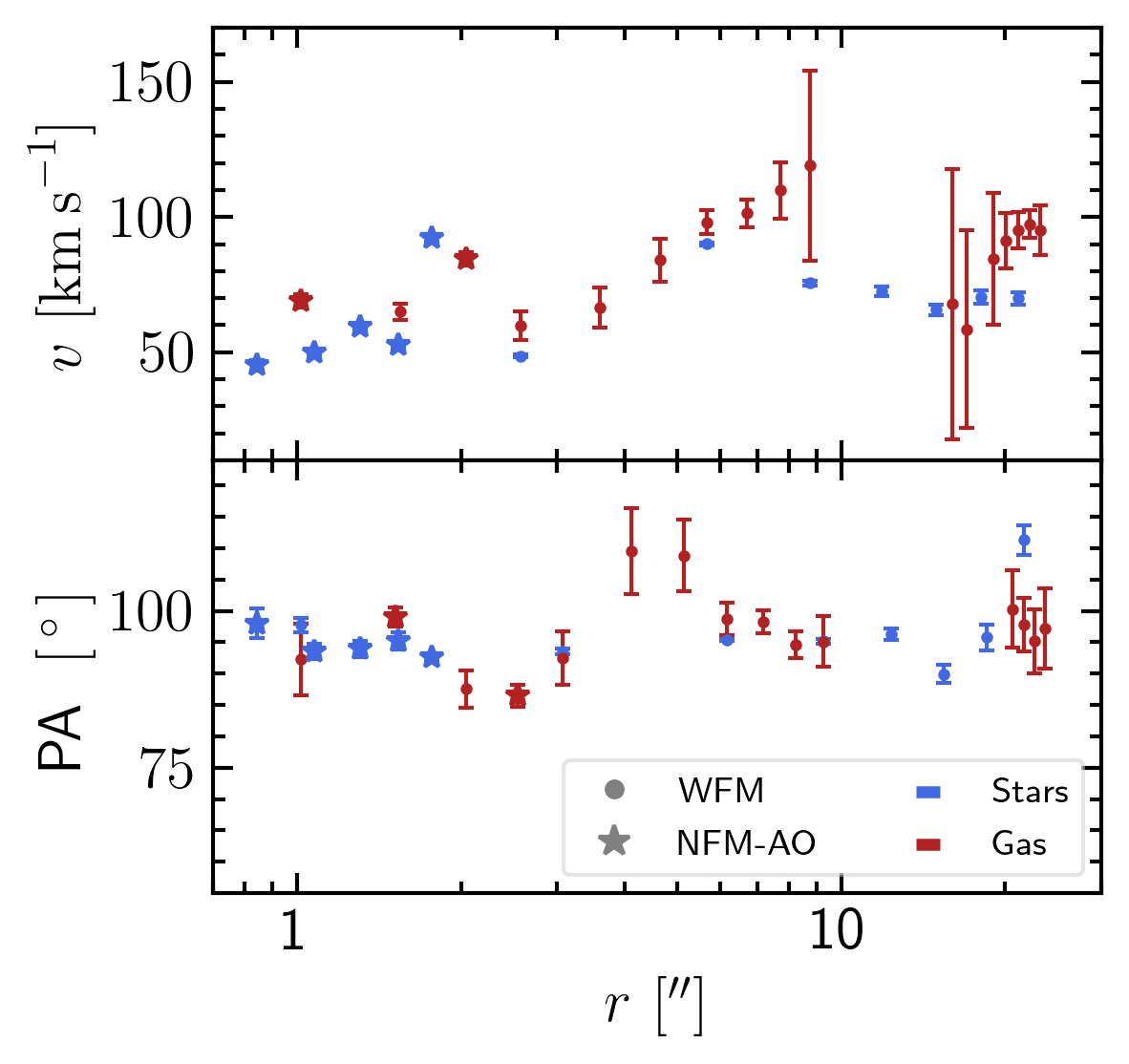} 
   \caption{Radial dependence of the rotating disk model parameters. Both model parameters of the galaxy stellar (blue) and ionized gas (blue) velocity field are shown. The data points acquired from the MUSE NFM-AO observations are shown as stars and have errors smaller than the size of the symbol.
   The data points from MUSE WFM are shown as errorbars.
   From $0\farcs8$ to $25\arcsec$ the PA of the disk-like rotation is constant and matches between ionized gas and stars. For both, the rotational velocity increases from the center up to $\sim$6$\arcsec$ before it flattens out.}
              \label{fig:rotation_curve}%
\end{figure}

Compared to the stellar rotation, the ionized gas rotates significantly faster than the stellar component as shown in Fig.~\ref{fig:vstar_vgas}. Since the stars experience a pressure support from their velocity dispersion, their motion in a disk system is not described by circular orbits. The higher rotational velocity of the ionized gas component is a reflection of this effect, the so-called asymmetric drift.

We model the ionized gas velocity on the native MUSE pixel sampling since the velocities measured from the high S/N emission lines are well constrained. Since the high velocity gas in the center is clearly not associated with the rotating disk of the host, we use a circular aperture to mask the pixels in the innermost $0\farcs5$. 
Furthermore, the ionized gas velocity field captured by MUSE WFM exhibits an arc-shaped stream perpendicular to the galaxy disk that stretches from 1.3-4$\,{\rm kpc}$. \cite{Powell:2018} suggested that this ‘bloom’ was first ejected by the high star formation surface density in Mrk~1044’s CNE. Since it does not co-rotate with the galaxy disk, we also exclude this feature from the model.

Apart from the two features that we manually excluded, the ionized gas velocity field appears quiescent, indicating an undisturbed rotational motion. 
The bottom right panels of Fig.~\ref{fig:kin} show that on galaxy scales the ionized gas is surprisingly well described by the disk rotation. Only within the innermost $1\,{\rm kpc}$ we detect spatially extended structures in the velocity residuals that coincide with the location of the star forming CNE.
We highlight these extended structures in Fig.~\ref{fig:arc}, where the ionized gas velocity residuals show both receding and approaching components with an spiral-like shape. We select two prominent patches which were obtained from smoothing the map with a Gaussian kernel of $3 \, {\rm px}$ and imposing a minimum velocity of $\pm 3\, {\rm km\:s}^{-1}$. The contours of the smoothed map together with the original map are shown in Fig.~\ref{fig:arc}. The size and orientation of the clumps coincide with the direction and extent of the spiral pattern of the dust absorption seen in the HST WFC3 image (see left panel of Fig.~\ref{fig:HST_image}). The arcs reach from $400\,{\rm pc}$ to the $150\,{\rm pc}$ towards the central region as highlighted in Fig.~\ref{fig:arc}. Both regions are ionized by star formation and have symmetric line shapes which indicate that the velocity residuals are genuine rather than artefacts from fitting asymmetric emission lines.
The approaching region has a median velocity offset of $\Delta v_r = 5.9 \,{\rm km\:s}^{-1}$ and the receding region $\Delta v_r = -7.0 \,{\rm km\:s}^{-1}$. Clump-like structures are typical signatures of radial streaming motions within the disc, likely due to slow inflow in the disc plane.

Fig.~\ref{fig:rotation_curve} shows the radial dependence of the kinematic parameters measured with the rotating disk model. As expected for an inclined rotating disk, the rotational velocity increases with increasing distance from the center up to a maximum value of $v_{\rm gas} \approx 100 \,{\rm km\:s}^{-1}$ and $v_\star \approx 80 \,{\rm km\:s}^{-1}$ respectively before the rotation curve flattens. Also the model confirms that the ionized gas rotates faster at all radii which is expected due to the asymmetric drift (see Fig.~\ref{fig:vstar_vgas}).

\cite{Gadotti:2020} report that the nuclear discs have larger rotational support as compared to the underlying main disc. For Mrk~1044 we recognize a velocity jump around $\sim 200\,{\rm pc}$ where both the ionized gas and the stellar rotation appear to rotate faster in the CNE.
However, the seeing-limited resolution of the MUSE WFM data of FWHM=$1\farcs03$, the rotation velocity in the innermost $\sim 330\,{\rm pc}$ is underestimated due to beam smearing. Only MUSE NFM-AO allows to resolve the stellar motion on scales of the CNE, but doesn't have the coverage to compare with larger scales. Thus, the velocity jump could also be caused by the different resolution of MUSE WFM and NFM-AO.
Within the uncertainties, the PA of the rotational axis is constant across the whole galaxy. The median values of ${\rm PA}_\star = 95.2 \pm 2.4 \,^\circ$ and ${\rm PA}_{\rm gas} = 99.0 \pm 16.2 \,^\circ$ indicate that both stellar and gas rotation are aligned.
Altogether, Mrk~1044’s kinematics are quiescent. The smooth radial behaviour of the kinematic parameters indicates that Mrk~1044 does not currently experience major interactions with a companion or external disturbances.

\section{Discussion}

Despite its super-Eddington accretion rate, Mrk~1044 shows an unperturbed, rotation dominated velocity field. Moreover, there is no close companion visible in continuum light that could trigger a nuclear starbursts or AGN activity. Alternatively, intrinsic processes could be responsible for the enhanced star formation activity near the nucleus.

\subsection{Star Formation as potential driver for the AGN fuelling}

\begin{figure*}
   \centering
   \includegraphics[width=.7\textwidth]{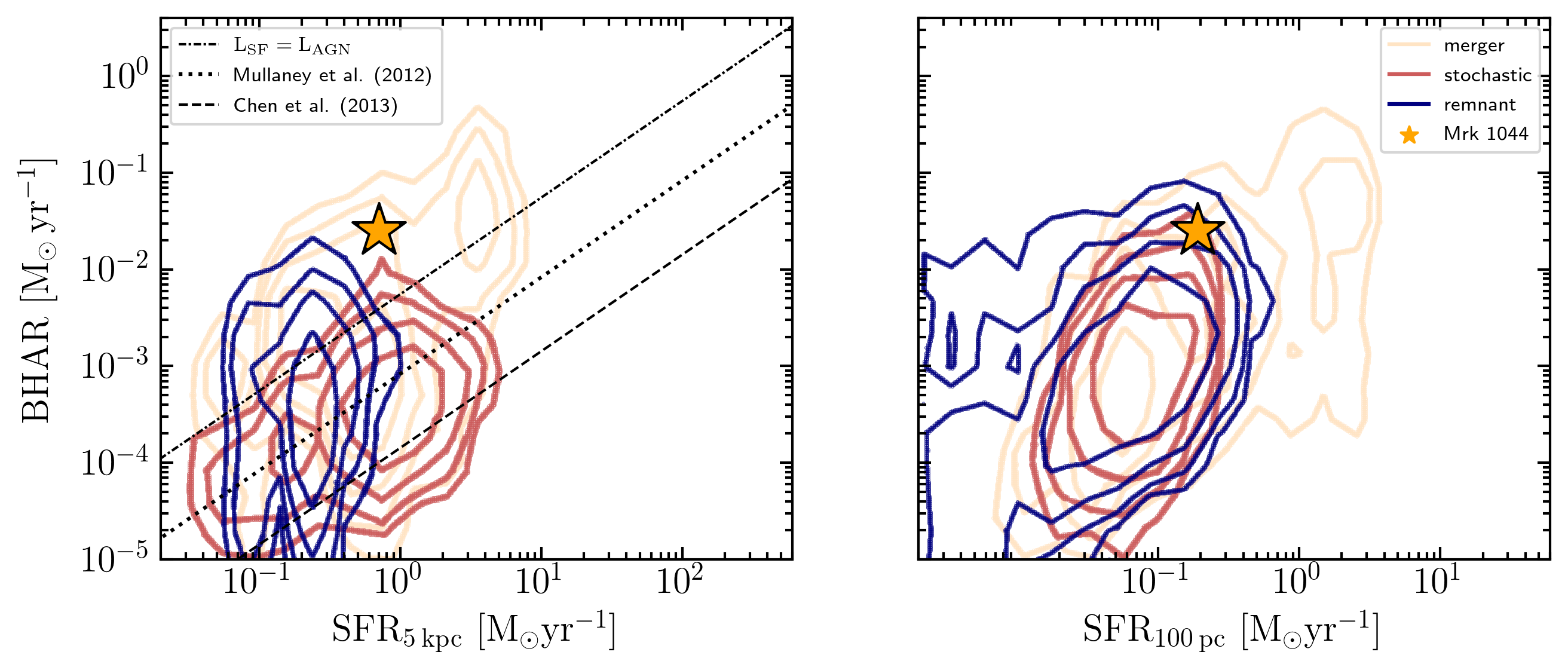} 
   \caption{Comparison of Mrk~1044’s BHAR and SFR on different spatial scales with theoretical predictions. The contours are taken from \cite{Volonteri:2015} and describe the BHAR during merger (blue), stochastic (red) and remnant (yellow) phase respectively.
   The lines represent the relations between galaxy-scale SFR and BHAR from \citealt{Mullaney:2012} (mass-selected galaxies, dashed line) and \citealt{Chen:2013} (star forming galaxies, dashed). The dash-dotted line marks the line that separates AGN and SF dominated regions.
   On galaxy scales Mrk~1044 has a high star formation rate that compares to that of merging galaxies. In the center, the high SFR of Mrk~1044’s CNE follows the predicted correlation with BHAR.
   }
              \label{fig:Volonteri+15}%
\end{figure*}

Contradicting results regarding the connection between star formation and AGN have been reported in the literature. While some authors claim a correlation between global SFR and AGN luminosity \citep[e.g.][]{Bonfield:2011,Rosario:2012, Chen:2013,Xu:2015,Lanzuisi:2017,Stemo:2020} others find only weak or no correlation \citep{Baum:2010,Rosario:2013,Azadi:2015,Stanley:2015,Stanley:2017}. This discrepancy can partially be explained by the different timescales of SF and AGN variability and the fact that the time-delay to trigger SF is typically longer than the AGN duty cycle \citep{Silk:2013, Zubovas:2013,Harrison:2021}. In addition, different spatial scales over which galaxy integrated properties are measured ($\sim$10$\,$kpc) compared to the BH accretion disk (<1$\,$pc) add further scatter to the correlation.

The importance of the distance on which AGN affect their hosts has been pointed out by \cite{Volonteri:2015} who used a suite of hydrodynamical simulations to compare the BHAR with the galaxy SFR on different spatial scales. We show their results together with Mrk~1044’s position in Fig.~\ref{fig:Volonteri+15}.
In their simulations, \cite{Volonteri:2015} describe galaxies as ioslated (‘stochastic’ phase), during a ‘merger’ that lasts $\sim 0.2 - 0.3 \,{\rm Gyr}$ or in the ‘remnant’ phase.
Their simulations predict that, in contrast to works from \cite{Mullaney:2012} and \cite{Chen:2013}, the galaxy-wide SFR is uncorrelated with BHAR. Instead, only the nuclear ($\sim$100$\,{\rm pc}$) SFR correlates with BHAR.
Mrk~1044 is in-line with this prediction. We have shown that on galaxy scales, the star forming CNE contributes a significant part of the total SFR which pushes Mrk~1044’s into the merger regime. On nuclear scales, however, the CNE in Mrk~1044 follows the ${\rm SFR}_{\rm 100\: pc}$-BHAR correlation. 
\cite{Volonteri:2015} further find that nuclear SFR and BHAR variations have similar amplitudes and both vary on similar timescales. 

These results are supported by starbursts observed in the inner regions of nearby Seyfert 1s where the starburst intensities increase with AGN luminosity and Eddington ratio \citep{Imanishi:2004,Davies:2007,Watabe:2008,Diamond-Stanic:2012}.
In this regard, Mrk~1044 represents an extreme example where the high star formation rate surface density coincides with a phase of strong nuclear activity. The critical question is whether both processes are physically connected?

\subsubsection*{SFR vs. BHAR}

Since a substantial fraction of Mrk~1044’s galaxy-wide star formation is concentrated in a circumnuclear ellipse, we want to find out whether it can explain the enhanced BHAR. Assuming that in Mrk~1044’s star forming CNE massive stars quickly return their material to the ISM through stellar winds and SNe, we can compare the mass fluxes from star formation and BH accretion.

H$\alpha$ nebular emission arises from the recombination of gas ionized by O- and early-type B-stars with M$_\star \gtrsim$ 20$\,{\rm M}_\odot$ \citep[][]{Peters:2017}. Due to their short lifetime, the presence of H$\alpha$ emission implies ongoing star formation within the last $5\,{\rm Myr}$.
We assume that the lifetime of the massive stars is short compared to the total duration of the present day star formation activity, i.e. they instantaneously return their material to the ISM except for a remnant mass $M_r$. We adopt the metallicity dependent compact remnant mass function from \citet[][eq. 5,7 and 9]{Fryer:2012}. Due to the vast abundance of molecular hydrogen in the host galaxy ($4 \times 10^8 \,{\rm M}_\odot$, \citealt {Bertram:2007}), we assume a constant SFR that equals the present-day value of $0.19 \,\rm{M_\odot yr}^{-1}$ estimated in Sect.~\ref{Sect:SFR}. We can express the rate of enriched material that fading stars return into the ISM as
\begin{align}
\label{eq:dotMSNe}
\dot{M}_{\rm SNe} = \int _{m_{min}} ^{m_{max}} [m-M_r] \times {\rm SFR} \times \xi(m) dm
\end{align}
where $\xi(m)$ is the initial mass function (IMF) and ${\rm SFR} \times \xi(m)$ corresponds to the birth rate of stars with mass $m$. We chose a minimum mass of $m_{min}=8\,{\rm M_\odot}$ for stars that enrich the ISM on the relevant timescale of $\sim 10\, {\rm Myr}$. Further, we assume a default Starburst99 IMF \citep{Leitherer:1999} from 0.1$\,{\rm M}_\odot$ to 100$\,{\rm M}_\odot$ and metal-rich gas ($Z/Z_\odot$=2) to estimate the mass ejection rate from stellar winds with Eq.~\ref{eq:dotMSNe}
\begin{align}
\dot{M}_{\rm SNe} = 3.6 \times 10^{-2} \,\rm{M}_\odot {\rm yr}^{-1}
\end{align}

Using Mrk~1044’s bolometric luminosity \mbox{ $L_{bol} = (1.4 \pm 0.2) \times 10^{44}\,{\rm erg\:s}^{-1} $} reported in \cite{Husemann:2022} and a radiative efficiency of $\eta = 0.1$, we estimate the black hole accretion rate (BHAR)
\begin{align}
    \dot{M}_\bullet = \frac{L_{bol}}{\eta c^2} = 2.5 \times 10^{-2} \,{\rm M}_\odot {\rm yr}^{-1}
\end{align}

Under the above assumptions the present-day net mass flux from stellar winds and SNe in the CNE exceeds the current BHAR by a factor of 1.5. Since we have not made a geometric assumption, comparing both quantities requires that all stellar ejecta remain within the CNE and are channeled towards the accretion disk which may not be the case in reality. We also note that choosing a higher $\eta$ that is typical of high accretion rates \citep{Ohsuga:2005,McKinney:2015} reduces the BHAR. Nonetheless, our simple model supports the scenario where the star formation in the circumnuclear ellipse drives the high accretion rate of Mrk~1044.

\subsubsection*{Lifetime of the CNE star formation}

The abundance of dense molecular gas from which the stars are formed correlates with the BHAR \citep{Izumi:2016, Shanggan:2020}. Theoretical models suggests a physical connection between molecular gas abundance and BHAR on scales of a few 100$\,{\rm pc}$ down to the accretion disc, independent from galaxy-scale processes \citep{Umemura:1997,Thompson:2005,Kumar:2010}.

We can estimate the maximum duration of the BH growth at the current rate by assuming a closed box model. The mass evolution of the gaseous disk with the cold gas mass $M_{\rm H_2}$ is given by
\begin{align}
\dot{M_{\rm H_2}} =  \dot{M}_{\rm SNe} -\dot{M}_\bullet -{\rm SFR}
\end{align}
Assuming that the cold gas content $M_{\rm H_2} = 4\times 10^8 \,{\rm M}_\odot$ \citep{Bertram:2007} will be entirely accreted onto the BH, the maximum duration of the gas accretion is
\begin{align}
\frac{M_{\rm H_2}}{\dot{M}_{\rm H_2}} &= \frac{M_{\rm H_2}}{\dot{M}_{\rm SNe} -\dot{M}_\bullet -{\rm SFR}} \\
&= 1.5 \,{\rm Gyr}
\end{align}

With this simplistic estimate, the central BH of Mrk~1044 will have consumed its entire pristine cold gas reservoir after 1.7$\,{\rm Gyr}$ with a final mass $M_\bullet = 4\times 10^8$ that equals the present-day HI mass. However, we already see a spatially resolved outflowing component in the ionized gas kinematics (see Fig.~\ref{fig:eline_maps}). Neglecting in- and outflows from the CNE might therefore not be a robust assumption. Using the BH mass - extended narrow line region size relation from \citep{Husemann:2022}, the expected timescale for an AGN-driven outflow to propagate through the CNE is $\sim 10^{4.5}\,{\rm yr}$, well below the time that the BH requires for significant mass growth. 
Furthermore, it is questionable whether the strong AGN phase will affect the host galaxy star formation activity through AGN feedback. Due to the proximity of the CNE near the active nucleus, the energy input irrespective of the coupling mechanism would first affect the CNE star formation before it would propagate through the entire host galaxy. The presence of ongoing star formation so close to the nucleus means that Mrk~1044’s BH fuelling is a recent event.
A detailed discussion on Mrk~1044’s AGN-driven outflow and its future impact on the host galaxy SFR will be presented in Winkel et al. (in prep.) 

\subsection{Chemical Enrichment of the CNE}

\begin{figure}
   \centering
   \includegraphics[width=.7\columnwidth]{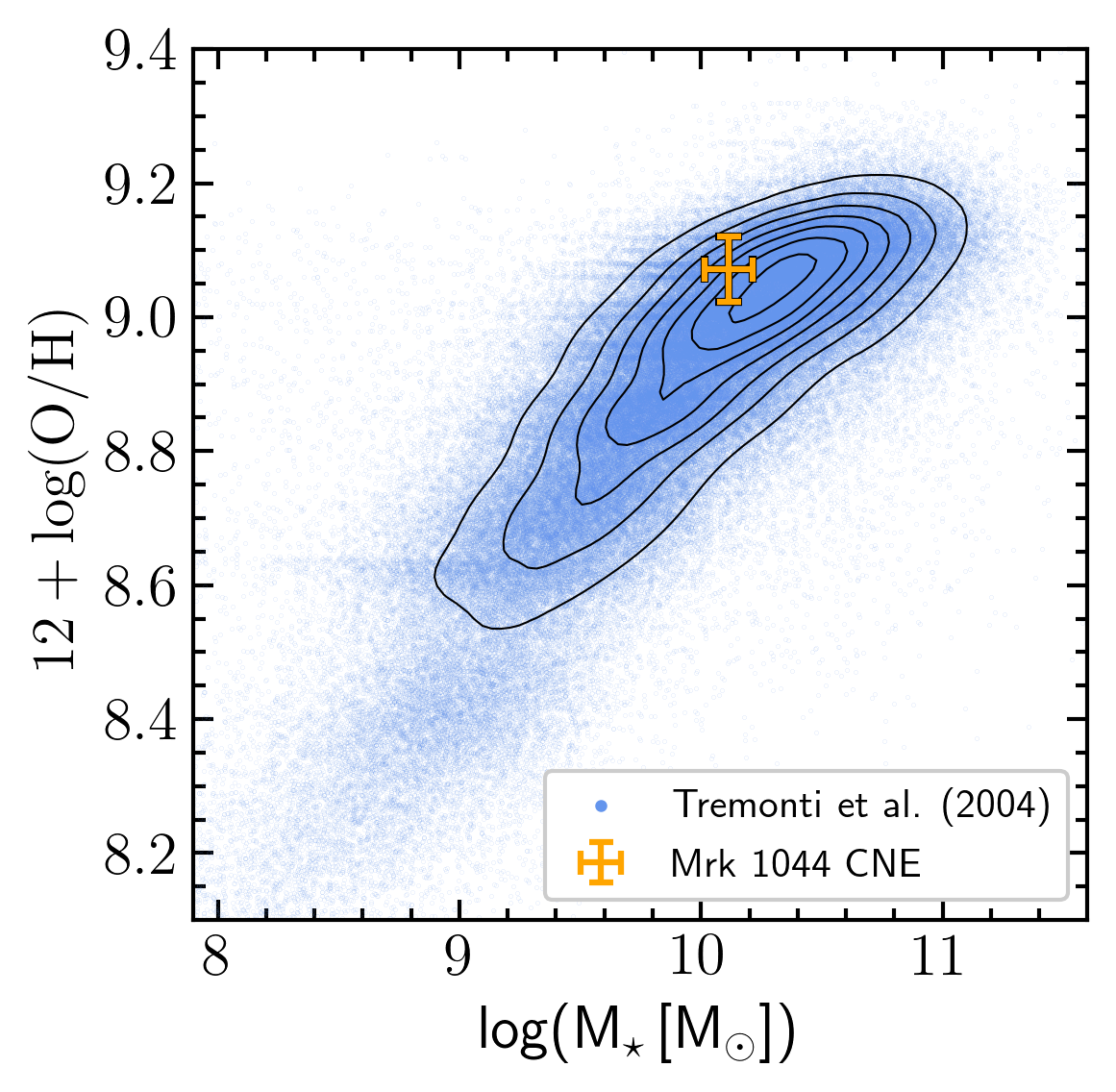} 
   \caption{Mrk~1044’s gas-phase metallicity relative to the mass-metallicity relation from \cite{Tremonti:2004}. Compared to star forming galaxies with similar stellar mass, Mrk~1044’s CNE has higher metallicity by $0.11\,{\rm dex}$.}
              \label{fig:mass-metallicity}%
\end{figure}

Using UV spectra obtained with HST Imaging Spectrometer
\cite{Fields:2005a} have identified two outflowing absorbing systems that escape from Mrk~1044’s center. From the column densities of \ion{O}{iv}, \ion{N}{v} and HI measured with the Far Ultraviolet Spectroscopic Explorer \cite{Fields:2005b} estimated their metallicity to be at least five times solar. Such high metal abundances are difficult to explain in a galaxy evolution context.

Fig.~\ref{fig:mass-metallicity} shows that, compared to the overall star forming galaxy population Mrk~1044’s star forming CNE has a gas-phase metallicity that is higher-than-average by $0.11\,{\rm dex}$. 
This result is in agreement with the unperturbed velocity field, as we do not see signs of an external gas reservoir that could fuel Mrk~1044’s center with fresh metal-poor gas. Instead, the high SFR suggests that Mrk~1044 is currently diverging from the mass-metallicity relation. The CNE size and its enrichment state are well in agreement with the TIMER studies of nuclear discs and nuclear rings \citep{Bittner:2020}. With an approximate bar size of $\sim 5\,{\rm kpc}$ bar (see Fig.~\ref{fig:HST_image}), Mrk~1044 fits well into the correlation between the nuclear discs size and bar size \citep{Gadotti:2020}, which suggests a secular bar-driven formation of the CNE.

\subsection{On the black hole feeding mechanism}

We have shown above that the mass ejection rate from dying stars in Mrk~1044’s circumnuclear star formation region is comparable to the current BHAR. Furthermore, the high metal abundances in the CNE and the accretion disk suggests that the gas has the same origin. To provide both the ongoing SF and the AGN with material, there must be efficient gas channelling from galaxy scales towards the center.
In the following we discuss the secular host galaxy processes that could trigger the radial gas migration.\\

\par 

Since the ionized gas co-rotates with the galaxy disk, the gas transport towards the center must be accompanied by a significant loss of angular momentum. One ingredient to achieve this is to increase the turbulence in the ISM which allows the gas to dissipate its angular momentum. There have been several disk-internal processes suggested that induce turbulence. For example, SN explosions and stellar winds deposit energy into to a pressure-supported circumnuclear gas disk. The higher turbulence increases the effective kinetic viscosity, i.e. enables a more efficient angular momentum transfer \citep{Kawakatu:2008, Wutschik:2013}. For Mrk~1044, the high SFR in the circumnuclear ellipse could be a reflection of this process.

Another process has been suggested for collisional disks where the cold gas could radially migrate towards the galaxy center via chaotic cold accretion \citep[CCA, e.g.][]{Gaspari:2019}. The circular motion is perturbed by recurrent collisions between cold gas cells which reduces their angular momentum and thus enable radial mass transport towards the center. In this way CCA can rapidly boost inflow rates from the galaxy meso- $(\sim 100\,{\rm pc})$ down to the micro-scale (<1$\,{\rm pc}$, \citealt{Gaspari:2017_cca}).
Independent of how it is injected into the rotating disk, turbulence allows the gas to dissipate its angular momentum and
radially migrate towards the galaxy center. The enhanced star formation seen in Mrk~1044’s CNE could either be a result of gravitational instabilities within the self-gravitating eccentric disk or vice-versa. \\

\par

Other astrophysical processes may further boost the angular momentum transport.
The strong eccentricity of Mrk~1044’s CNE suggests that its formation is driven by a galaxy scale process. 
Furthermore, Mrk~1044’s host morphology is not that of a bona fide barred spiral. As visible in the left panels of Fig.~\ref{fig:HST_image}, the inner $\sim$500$\,{\rm pc}$ are dominated by dusty spiral arms and are slightly elongated from the south-east to the north-east. 
The spiral arms are surrounded by a tenuous bar with a semi-major axis of $\sim$2$\,{\rm kpc}$ that in contrast stretches in the east-west direction. 
On the largest scales up to $\sim$15$\,{\rm kpc}$, the galaxy outskirts exhibit once more a spiral structure that is elongated in the north-south direction. 
Due to its peculiar host morphology, the gas migration towards Mrk~1044’s central engine could be driven by angular momentum transport through gravitational torques \citep{Shlosman:1990}. The minor axis of the CNE coincides with the orientation of the bar and is therefore likely a resonant ring that typically forms at the inner Lindblad resonances (ILR). At that location of the ILR, the gas-flow towards the center is slowed down and it accumulates in a reservoir \citep{Combes:1996, Comeron:2010}. This is is supported by the enhanced SF within Mrk~1044’s circumnuclear ellipse which is likely the reflection of a gas repository. 

\par

At smaller distances than 100$\,{\rm pc}$ from the center, large-scale torques become inefficient and disk instabilities are required to transport gas down to scales of the accretion disk \citep["bars within bars",][]{Shlosman:1989}. Hydrodynamic simulations from \cite{Hopkins:2010} show that this process would explain the high star formation rate through the formation of circumnuclear molecular gas clumps. This means that gas accretion and SF are self-regulating and can coexist in an inhomogeneous, gravitationally unstable disk \citep{Vollmer:2018, Schartmann:2018}.
In this scenario, the gas may only be partially converted into stars since the advection timescale is shorter than the star formation timescale \citep{Thompson:2005}. Therefore, the radial gas transport allows for efficient channelling to the accretion disk, which would explain Mrk~1044’s high accretion rate. As the motion of the stellar component is driven by the gravitational potential alone, the angular momentum transport only affects the gas component. 

For Mrk~1044, we see a possible signature of this process in the spiral trails in the ionized gas velocity field (Fig.~\ref{fig:arc}). The majority of galaxy spiral arms are trailing \cite[e.g.][]{Pasha:1982, Puerari:1992} which can be explained with swing amplification of density waves that direct inwards \citep{Toomre:1981, Athanassoula:2009}. Thus, the deviation from circular rotation likely traces the radial gas migration towards Mrk~1044’s galaxy center. On the other hand, the ionized gas contributes a small fraction of the supply for the SMBH compared to the cold molecular gas \citep[e.g.][]{Gaspari:2017_cca}. While the ionized gas is accelerated by momentum injection through radiation and winds form SNe, the dense inner regions of molecular clouds might be shielded from both and its motion remain unperturbed. 
However, in a CCA scenario, the molecular gas is expected to inherit part of the galaxy-scale kinematics via the top-down condensation, hence leading to tighter phase correlations \citep[][]{Gaspari:2018}.
Overall, we require high angular resolution of the molecular gas kinematics in Mrk~1044’s central region to understand the detailed gas feeding towards the accretion disk, in particular to unveil the cooler phases.

\section{Summary and conclusions}

In this work we have presented new MUSE NFM-AO observations of the nearby unobscured NLS1 galaxy Mrk~1044. The deblending of the AGN from the host emission enabled us to extract the host galaxy properties with unprecedented resolution.
We have mapped the host galaxy stellar velocity field and the emission line properties in Mrk~1044’s central ${\rm kpc}$. Furthermore, we have used a kinematic model to identify structures that deviate from a quiescent rotating disk. Our key results are the following.

\begin{itemize}

\item{A large fraction (27\%) of the galaxy-wide star formation is concentrated in a compact star forming circumnuclear ellipse (CNE) with $\Sigma_{\rm{SFR}} = 0.16 \pm 0.04 \, \rm{M}_\odot \rm{yr}^{-1}\rm{kpc}^{-1}$, a semi-minor axis of \mbox{$306 \pm 43\,\rm{pc}$} and an axis ratio of 2.45. Within the CNE the optical emission line ratios are entirely consistent with excitation by star formation. Surprisingly, we do not find signs of extended photo-ionized gas emission, despite Mrk~1044’s high accretion rate.}

\item{From 30$\,{\rm kpc}$ down to $\sim$100$\,{\rm pc}$ both stellar and ionized gas follow the quiescent kinematic pattern of a rotating thin disk. In the velocity fields, there are no signs of a major interactions with a companion.}

\item{Within the compact CNE, the ionized gas exhibits clumpy streams with $\sim 100\,{\rm pc}$ extent that deviate from rotation with the disk, potentially contributing to the SMBH feeding. Their projected line-of-sight velocity is of the order of $10 \,{\rm km\:s}^{-1}$.}

\item{The gas-phase metallicity within the star forming CNE is above the expected value from the mass-metallicity relation.}

\item{The estimated mass ejection rate from massive stars exceeds the current BHAR.}

\end{itemize}

Our results suggest that the circumnuclear star formation is connected to the high accretion rate of Mrk~1044’s central black hole.
It remains unclear, however, if the star formation drives Mrk~1044’s black hole accretion or if both share an underlying process that triggers them at the same time.
Very long baseline interferometric observations of Mrk~1044’s nuclear region have been conducted recently with the Atacama Large Millimeter/submillimeter Array (ALMA). The information on the molecular gas abundance and distribution will help to understand the fuelling of Mrk~1044’s central engine. However, to constrain the initial process driving the rapid black hole growth in NLS1, a comparative study of nearby BLS1s and NLS1s with high spatial resolution is required.

\begin{acknowledgements}
    We thank the anonymous referee for suggestions and comments that helped to improve the presentation of this work.
    NW would like to thank Eric C. Rohr for frequent discussions. This work was funded by the DFG under grant HU 1777/3-1. 
    NW, BH and ISP acknowledge travel support from the DAAD under grant 57509925 and the hospitality of MS at the University of Manitoba where large parts of this paper were written. 
    BH was partially supported by the DFG grant GE625/17-1. TAD acknowledges support from the UK Science and Technology Facilities Council through grants ST/S00033X/1 and ST/W000830/1. 
    V.N.B. gratefully acknowledges assistance from National Science Foundation (NSF) Research at Undergraduate Institutions (RUI) grant AST-1909297. Note that findings and conclusions do not necessarily represent views of the NSF. CO and MS are supported by the Natural Sciences and Engineering Research Council (NSERC) of Canada. 
    MG acknowledges partial support by NASA Chandra GO9-20114X and {\it HST} GO-15890.020/023-A, and the {\it BlackHoleWeather} program.
    The Science and Technology Facilities Council is acknowledged by JN for support through the Consolidated Grant Cosmology and Astrophysics at Portsmouth, ST/S000550/1.
    The work of MS was supported in part by the University of Manitoba Faculty of Science Graduate Fellowship (Cangene Award), and by the University of Manitoba Graduate Enhancement of Tri-Council Stipends (GETS) program.
\end{acknowledgements}

%
%

\bibliographystyle{aa}
\bibliography{references}

\end{document}